\documentclass{article}

\usepackage{bm}
\usepackage{authblk}
\usepackage{graphicx}
\usepackage{cancel}
\usepackage{caption}
\usepackage{subcaption}
\usepackage{amsmath}
\usepackage{amssymb}
\usepackage{xcolor}
\usepackage[title,toc,titletoc,page]{appendix}
\usepackage{hyperref}
\usepackage{soul}
\usepackage{tikz}
\usetikzlibrary{shapes.geometric, arrows, positioning}

\hypersetup{
    bookmarksnumbered=true, 
    bookmarksopen=true,     
    pdfpagemode=UseOutlines 
}

\tikzstyle{startstop} = [rectangle, rounded corners, minimum width=0.2cm, minimum height=0.2cm, text centered, draw=black, fill=red!30]
\tikzstyle{process} = [rectangle, minimum width=0.3cm, minimum height=0.1cm, text centered, draw=black, fill=orange!30]
\tikzstyle{pythonprocess} = [rectangle, minimum width=0.3cm, minimum height=0.1cm, text centered, draw=black, fill=blue!30]
\tikzstyle{arrow} = [thick,->,>=stealth]

\title{{Hammering at the entropy}: A \textsc{generic}-guided approach to learning polymeric rheological constitutive equations using PINNs.}
\author[1]{David Nieto Simavilla} 
\author[2,3]{Andrea Bonfanti}
\author[4]{Imanol García de Beristain} 
\author[5]{Pep Español}
\author[6,7,8]{Marco Ellero}

\affil[1]{Dept. Energía y Combustibles, Escuela Técnica Superior de Ingenieros de Minas y Energia, Universidad Politécnica de Madrid}
\affil[2]{BMW Group, Digital Campus Munich}
\affil[3]{University of the Basque Country (UPV/EHU)}
\affil[4]{Applied Mathematics Department, Engineering School of Bilbao,
University of the Basque Country (UPV/EHU)}
\affil[5]{Dept. Física Fundamental, Universidad Nacional de Educación a Distancia}
\affil[6]{Basque Center for Applied Mathematics (BCAM)}
\affil[7]{IKERBASQUE, Basque Foundation for Science}
\affil[9]{Zienkiewicz Centor for Computational Engineering (ZCCE), Swansea University}

\begin{document}

\maketitle

\begin{abstract}
We present a versatile framework that employs Physics-Informed Neural Networks (PINNs) to discover the entropic contribution that leads to the constitutive equation for the extra-stress in rheological models of polymer solutions. In this framework the training of the Neural Network is guided by an evolution equation for the conformation tensor which is \textsc{GENERIC}-compliant. We compare two training methodologies for the data-driven PINN constitutive models: one trained on data from the analytical solution of the Oldroyd-B model under steady-state rheometric flows (PINN-rheometric), and another trained on in-silico data generated from complex flow CFD simulations around a cylinder that use the Oldroyd-B model (PINN-complex). The capacity of the PINN models to provide good predictions are evaluated by comparison with CFD simulations using the underlying Oldroyd-B model as a reference. Both models are capable of predicting flow behavior in transient and complex conditions; however, the PINN-complex model, trained on a broader range of mixed flow data, outperforms the PINN-rheometric model in complex flow scenarios.  The geometry agnostic character of our methodology allows us to apply the learned PINN models to flows with different topologies than the ones used for training.

\end{abstract}

\tableofcontents

\section{Introduction}
Rheology  aims to  predict the  complex flows  of viscoelastic  fluids
using balance  equations, equations  of state (EOS),  
and constitutive
equations. These  equations can  be solved  once initial  and boundary
conditions are  provided. While  balance equations are  universal, EOS and  constitutive equations  are system specific. {Therefore,
models for the EOS, 
relating temperature and
pressure  to  mass and  energy  densities,  and for  the  constitutive
equations, relating stress  and microstructural conformation tensors are required.}   The exact
functional    form   of    these   models    is   generally    unknown \cite{Schieber_book, Bird1987}.

Traditionally, physical models are postulated based on physical intuition and are characterized by a small number of parameters that, ideally, can identify physical constants and material properties \cite{Dyson2004}.   In particular, rheologists have dedicated huge efforts to find good models that describe complex flow properties and behavior \cite{oldroyd1950, HERRCHEN1997, thien77,Larson1988, Kroger2010}. Once a model is proposed, model parameters are typically determined through simple experiments (i.e., in rheology shear or extensional steady-state viscometric flows). These experiments are designed to fit or ``learn" the model’s parameters in order to achieve good predictions. 
Models with a  minimal number  of parameters enable their determination
with  a limited  amount of  data from  a few  simple flow  experiments
involving  shear  or  extensional steady-state  viscometric  flows.  A
fundamental   problem  in   rheology  is   whether  viscometric   flow
experiments are sufficient  to obtain the parameters of a  model to be
applied  in complex  flow  situations. A  good  physical model  should
describe a wide range of flow  situations beyond those used to fit the
parameters.  However,  there is no  guarantee that a  model performing
well in simple  flows will excel in complex flow  situations. When the model fails, it must either be improved or replaced by a new one that incorporates additional physical insight.  This  process often  entails identifying  key \emph{physical
  features},  such as  finite  extensibility,  constraints imposed  by
entanglements, etc.   that better capture the  system’s behavior. The
systematic process of  model refinement by  including new  physics, however, 
might be time-consuming,  and we may  also end up  with something  that looks  very
different  to a  \emph{good physical  model}  with a  small number  of
parameters \cite{Dyson2004}.

An alternative, is to use a model with a sufficiently large number of parameters to \emph{fit the elephant} \cite{Dyson2004}. We loose the physical intuition concerning the behaviour of the model but, provided that a sufficiently large amount of data is available,  we may automatize the entire process. Thanks to the universal approximation theorem \cite{universalapprox}, we may use a neural network (NN) to represent the functional form of a model with enough flexibility. The model is given by the composition of functions with very simple functional forms (like sigmoidal functions), resulting in a generic functional form depending on a large number of parameters. Using NN as the model, we may fit its parameters (train the network) in any region of the space of variables, be it the region of viscometric flows, or in arbitrary regions of complex flows. This allows us to simply upgrade the NN model with new information when we require predictions for flow simulations of increasing complexity. {Graphically, we hammer the function against data as a blacksmith would hammer a sheet of metal to shape it.} 

Traditional neural networks have been used in multiscale simulation methods to retain the microscopic molecular fidelity at the macroscale \cite{Lei2020}. Neural networks, while powerful function approximators, often disregard established constitutive modeling principles and thermodynamic constraints. As any interpolation method, NNs will give \emph{good predictions} in the vicinity of training points and will  fail extrapolating to regions far from their training data \cite{bonfanti2023,bonfanti2024}. Furthermore NNs typically require large data sets to achieve a decent training \cite{rackauckas2021}. In the context of rheology, data-driven approaches have been used to model nonlinear viscoelastic materials at small strains using neural networks (NNs), requiring only stress and strain paths for training. These NNs can be tuned to satisfy some physical feature (i.e., convexity of the learned functional) to facilitate the handling of large data sets and noisy stress data \cite{Rosenkranz2024}.

Several lines of research aim at leveraging the potential advantages of machine learning methodologies in the field of rheology. Young \emph{et al.} have employed scattering microstructure data to develop low dimensional constitutive models using a dimensionality reduction scheme \cite{Young2023}. Zhao and coworkers have employed yet a different machine learning methodology -- Gaussian Process Regression (GPR) -- to learn constitutive models from microscopic simulations under simple shear flows. These constitutive models can then be used in macroscopic simulations \cite{zhao2018,zhao2021}. However, an important limitation of these studies is that they set arbitrary constraints to the functional form of the learned rheological model. For example, setting the viscosity as function of shear rate alone \cite{zhao2018} or, having the microstructure description relying on a pre-defined FENE-P model to incorporate viscoelasticity \cite{zhao2021}, thus these approaches lack generality. A more general, yet similar in the \emph{ad hoc} choice of the functional form of the constitutive model is the work by Seryo \emph{et al.} \cite{Seryo2020}. GPR is also used to learn a constitutive model that introduces history-dependent viscoelasticity by considering the time derivative of the polymeric stress as a function of the flow velocity gradient and the stress \cite{Seryo2020}. 
However, as the model becomes more general the dimensionality also tends to grow. For example in Seryo \emph{et al.} work, the derivative of the stress $\frac{\partial \bm \tau}{\partial t}(\bm \nabla \bm v, \bm \tau)$ with a priory 6 independent components for 3D problems should be learned from the velocity gradient $\bm \nabla \bm v$ with 9 independent components and the symmetric stress $\bm \tau$ with additional 6 independent components (i.e., mapping 15-dimensions space into 6-dimensions space), making the approach difficult to apply in complex 3D flows. To accomplish it, a more effective, yet general, physics-informed dimensionality reduction is thus required. 


Another alternative is to utilize Physics Informed Neural Networks (PINNs) \cite{raissi2019physics}, a novel family of Neural Networks which are able to inherently satisfy kinematic, thermodynamic, and physical constraints \cite{Linka2023}.   
PINNs are neural networks that incorporate model equations, such as Partial Differential Equations (PDEs), directly into their structure. PINNs are currently employed to solve  forward and inverse PDE problems \cite{raissi2019physics}, fractional equations \cite{pang2019fpinns}, integral-differential equations \cite{yuan2022Apinn}, and stochastic PDEs \cite{zhang2019pinnstocastic}. This innovative functions act as a multi-task learning framework where the neural network simultaneously fits observed data and minimizes the residual of the selected  PDE \cite{Cuomo2022, Mahmou2022}. 
The introduction of governing equations in the loss function enables PINNs to offer a powerful framework for solving forward and inverse problems in fluid mechanics, where the solutions to Navier-Stokes equations that incorporate complex constitutive models are produced as predictions of the neural network \cite{Karniadakis2021}. However, most of the studies using PINNs have focused on predictions of specific flow simulations (i.e., using the network to solve Navier-Stokes equations) and then choose the right constitutive model from a selection of `known' analytical constitutive models rather than learning `unknown' constitutive models \cite{Meng2020, lin2023}. For example, Thakur \emph{et al.} used PINNs to select among a selection of viscoelastic constitutive models (Oldroyd-B, Giesekus or Linear PTT) and learn the stress field from a velocity field \cite{thakur2022}. 
In \cite{Saadat2022, Mahmou2022b} the authors have proposed the Rheology-Informed Neural Networks (RhINNs) for forward and inverse meta-modelling of complex fluids. The RhINNs are employed to solve the constitutive models with multiple Ordinary Differential Equations (ODEs) by proposing a penalisation based on a thixo-visco-elasto-plastic model (TVEP) for the stress, where a few model parameters are learnt. Again, the framework used to encapsulate the physics is over-restricting for general complex fluids.

Nevertheless, PINNs present a significant advantage over traditional data-driven models by ensuring that the neural network solutions adhere to fundamental physical principles. This feature is particularly beneficial in the context of rheology, where the complexity of the flow behavior demands models that respect the underlying physics while adapting to diverse and nonlinear phenomena. In this context a number of studies have employed PINNs to advance rheological modeling. An interesting use regarding the problem discussed above is to employ PINNs to model viscoelastic materials using deep neural networks to approximate rate-dependent and nonlinear constitutive relationships \cite{Xu2021}.
One way to impose a minimal set of thermodynamics-based constraints (i.e. First and Second Laws of Thermodynamics) on constitutive models, still retaining generality, is through the application of the so-called 
\textsc{generic} framework  (General Equations of Non-Equilibrium Reversible Irreversible Coupling) \cite{Ottinger_book}.
Hernandez \emph{et al.} proposed structure-preserving Neural Networks \cite{Hernandez2021}.
Zhang \emph{et al.} proposed a \textsc{generic} formalism Informed Neural Networks (GFINNs) that adhere to the symmetric degeneracy conditions of the \textsc{generic} formalism. GFINNs consist of two modules, each with two components, modeled by neural networks specifically designed to meet these conditions. This component-wise architecture allows flexible integration of physical information into the networks \cite{Zhang2022}.

In this work, we present a general approach for polymer solutions using PINNs to determine the polymeric entropy leading to the constitutive equation for the stress in rheological models. Instead of training PINNs to predict arbitrary solutions of specific flow simulations, we aim to leverage their universal approximator nature to capture the general functional relation between the eigenvalues of the conformation tensor $\bm c$ and the polymeric entropy for `a priori' unknown viscoelastic models.  The approach is not only  \textsc{generic}-compliant but also significantly reduces the problem dimensionality making model learning more efficient. In fact, it only requires learning a scalar state function (the entropy) as a function of the 2 or 3 eigenvalues (depending on space dimension) of the conformation tensor $\bm c$. 
We evaluate the traditional methods using limited regions of the available conformation space (i.e., limited data from steady-state rheometric flows) to establish rheological models that then can be used to predict properties and behavior of more complex flows. We propose two types of data sets to train our PINN models: a first one -in analogy to classical rheological calibrations - with steady-state rheometric flows (later denoted as ``PINN-rheometric"); and a second one, with data from steady-state solutions of complex flow around a cylinder (later denoted as ``PINN-complex"). We study the application of the PINN models to finite volume simulations of complex flows coupling the  learned models with an OpenFOAMs RheoTool solver \cite{PIMENTA2017,Alves2001}. We find that a PINN model trained in steady-state rheometric flows data can be used to produce reasonable predictions in moderate transient or complex flows. 
However, in order to reproduce complex flows more accurately, data retrieved beyond viscometric flows are required for training.


In order to evaluate the quality of the PINN model predictions, we analyze the relative errors of the entropy and the stress in the space of the eigenvalues of $\bm c$.
We generate in-silico conformation-tensor data according to analytical and accurate RheoTool-discretisations of an Oldroyd-B model in simple and complex flows. {Data are provided to the PINN for training, whereas its application to complex flow is agnostic of the underlying model used to generate them, therefore providing an optimal and controlled framework for fair numerical comparisons.} The PINN model presented here can effectively learn unknown forms of the polymeric entropy and integrate their \textsc{generic}-guided NN representation into RheoTool to perform data-driven flow simulations. The only data required to train these models are the conformation tensor and velocities fields in complex flow. 
These kind of datasets can be obtained, for example, in experiments through combined use of particle-image-velocimetry (PIV) and  flow-induced birefringence measurements \cite{Haward2021} or, in the case of multiscale applications, from indipendent mesoscale polymer computations \cite{Nieto2022,Nieto2023}. 
Thus, we aim  to leverage the potential of PINNs 
to provide rheologists with more effective, thermodynamics-guided ways to discover constitutive equations from data and, at the same time,  applying them directly to fluid mechanics simulations using CFD. 



\section{Methods}
\subsection{The \textsc{generic}-guided approach to model constitutive equations}

In  the present  work, we  are interested  in the  modeling of 
polymeric solutions. Within the  \textsc{generic} framework, a general
polymer model can be cast into a set of partial differential equations
including the  mass and  momentum balance  equations 
\begin{align}
\label{eq:eq_continuity}
&\boldsymbol{\nabla} \cdot \bm{v} = 0\\
\label{eq:eq_momentum}
&\rho \left(\frac{\partial \bm{v}}{\partial t} + \bm{v} \cdot \nabla \bm{v}\right) - \nabla \cdot \left(\frac{\eta_s}{2} (\nabla \bm{v} + \nabla \bm{v}^T ) \right)  = -\nabla p  + \nabla \cdot \boldsymbol{\tau}
\end{align}
where $\bm{v}$ is the velocity vector field, $p$ is the pressure, and 
$\bm \tau$ is the non-Newtonian extra-stress term
coupled  with an
evolution  equation  for  the  conformation  tensor  ${\bm  c}$,  that
represents    the   microstructure    generated   by    the   polymers
\cite{Ottinger_book}. For  a 
polymer solution undergoing affine deformation,  the conformation
tensor generally evolves according to \cite{vazquez2009b,Nieto2023}
\begin{align}
    &\partial_t {\bm c} = - {\bm v} \cdot {\bm \nabla} {\bm c} + {\bm  c}\!\cdot\!\bm {\kappa}
+ \bm {\kappa}^T\!\cdot\!{\bm  c} +\frac{2}{\lambda_{\rm p} n_{\rm p} k_BT}{\bm  c}\!\cdot\!\bm {\sigma} &\label{eq:const_c}
\end{align}
where $\lambda_{\rm p}$ is the polymeric characteristic relaxation time, $n_{\rm p}$ the polymer number density (i.e., the number of chains per unit volume), $k_{\rm B}$ is the  Boltzmann constant and $T$ the temperature.
The first three terms in Eq. (\ref{eq:const_c}) are reversible in nature,
describing the  kinematic evolution  of the conformation  tensor under
the influence of the velocity gradient $\boldsymbol{\kappa}=\nabla \bm{v}^T$. The last
term  describes   the general irreversible   evolution  of   the
conformation  tensor. 
This term  is  characterized  by 
the thermodynamic force
${\bm\sigma}({\bm c})$,  defined as  the  derivative of  the  polymeric  entropy density
$s_{\rm p}({\bm c})$ with respect to the conformation tensor $\bm c$, that is
\begin{align}
  \label{eq:sigma_def}
	\frac{\bm \sigma}{T}=\frac{\partial s_{\rm p}}{\partial \bm c}
\end{align}
Finally, the  momentum balance equation contains, in addition to the
solvent viscous stress, a polymeric stress given by   \cite{vazquez2009b}
\begin{align}
  \label{eq:tau}
\bm \tau = - 2 \bm c \cdot \bm \sigma  
\end{align}
which satisfies the dynamics-thermodynamics compatibility (i.e., consistency with the microstructural evolution given by Eq.(\ref{eq:const_c})). Therefore,  the knowledge  of the entropy  function directly provides the closure in  a thermodynamic-consistent constitutive equation for the polymeric suspension.

Since the entropy  is invariant under rotations,  it can only
depend on the invariants of the conformation tensor $\bm c$, that we choose to
be the eigenvalues    $c_1,c_2,c_3$ and therefore,
$s_{\rm p}({\bm c})=s_{\rm p}(c_1,c_2,c_3)$.   Observe   that  the   tensors  $\bm   c$  and
$\bm    \sigma$    commute   \cite{vazquez2009b}    and    diagonalize
simultaneously. This implies that we may write Eq.(\ref{eq:sigma_def}) as
\begin{align}\label{eq:eigen-sigma_def}
	\frac{\sigma_\alpha}{T}=\frac{\partial s_{\rm p}}{\partial c_\alpha}
\end{align}
where      $\sigma_\alpha$      ($\alpha=1,..,D$) are       the      eigenvalues      of
$\boldsymbol{\sigma}$. Because of the  large reduction of arguments of
the  entropy  due to  rotational  symmetry,  it proves  convenient  to
express the  dynamics in terms  of eigenvalues and  eigenvectors.  The
decomposition   of  Eq.(\ref{eq:const_c})   into  eigenvalues   and
eigenvectors (i.e., ${\bf c}=\sum_\alpha c_\alpha {\bm u}_\alpha {\bm u}_\alpha^{\rm T}$) leads to two coupled PDEs \cite{vazquez2009b}
\begin{align} 
	& 0 = \partial_t c_{\alpha} + v_j\partial_jc_{\alpha} -2c_{\alpha}\kappa_{\alpha\alpha} - \frac{1}{\lambda_{\rm p} n_{\rm p} k_{\rm B} T} c_\alpha  \sigma_\alpha \label{eq:refec} \\
&0 =  \partial_t \bm u_{\alpha} + v_j\partial_j\bm u_{\alpha} - \sum_{\beta} H_{\alpha\beta}^{\rm  } \bm u_{\beta} \label{eq:eu}
\end{align}
where $\kappa_{\alpha\alpha}={\bf u}_\alpha\cdot\boldsymbol{\kappa}\cdot {\bf u}_\alpha$ is the velocity gradient in the eigenbasis of the conformation tensor, and 
the anti-symmetric matrix $H_{\alpha\beta}^{\rm  }$ is given by: 
\begin{align}\label{eq:hmix}
H_{\alpha\beta}^{\rm  } = \frac{1}{c_\alpha-c_\beta}(c_\alpha \kappa_{\alpha\beta} + c_\beta \kappa_{\beta \alpha}) 
\end{align}
The kinematics of the flow in  Eq. \ref{eq:eu} has been used by the authors in \cite{Nieto2023}
to    establish    the     non-affine    characteristics    of    polymer
flow by introducing a mixed derivative of the Gordon-Schowalter type. 
This allows for the {unambiguous separation of reversible/irreversible terms in the dynamics}, enabling to split non-Newtonian effects related to non-affine deformation, with irreversible effects intrinsically associated to the polymeric entropy $s_p({\bf c})$, which is crucial 
to apply safely the model to arbitrary flows.
For simplicity, non-affine motion is not considered here as it will be the subject of future refinements of the methodology. Note that the    entropy    appears   only    in    Eq.
(\ref{eq:refec}) through $\sigma_\alpha$. 
Equation (\ref{eq:refec}) will be used below to construct the residuals and the loss function in the PINN model.

In this paper, we address the following problem: given that the
dynamics    of    the    polymer    solution    are    described    by
Eq. (\ref{eq:const_c})  or its equivalent, Eq.   (\ref{eq:refec}), and
that we  have explicit data  for the  fields ${\bm v}({\bf  r},t)$ and
${\bm c}({\bf  r},t)$, our  objective is to  develop a physics-informed neural network
(PINN) representation of  the specific entropy function $s_p({\bf  c})$.  This PINN
model should ensure that the  measured fields align with the governing
equation (\ref{eq:const_c})
dictated by \textsc{generic}.
It should be also noticed that our approach diverges from the usual application of PINNs where the
neural  networks  represent  the fields  ${\bm  v}({\bf r},t)$  and
${\bm c}({\bf r},t)$ directly  
\cite{Karniadakis2021}. In contrast, in this paper, we focus solely
on employing a  single neural network to model the  functional form of
the entropy. From this knowledge alone, $s_{\rm p}$ can be used to provide all the necessary stress predictions in CFD simulations (i.e., using RheoTool). 


\subsection{The neural network}
\label{sec:loss}
\begin{figure}[ht!]
  \centering
  \includegraphics[width=1.0\linewidth]{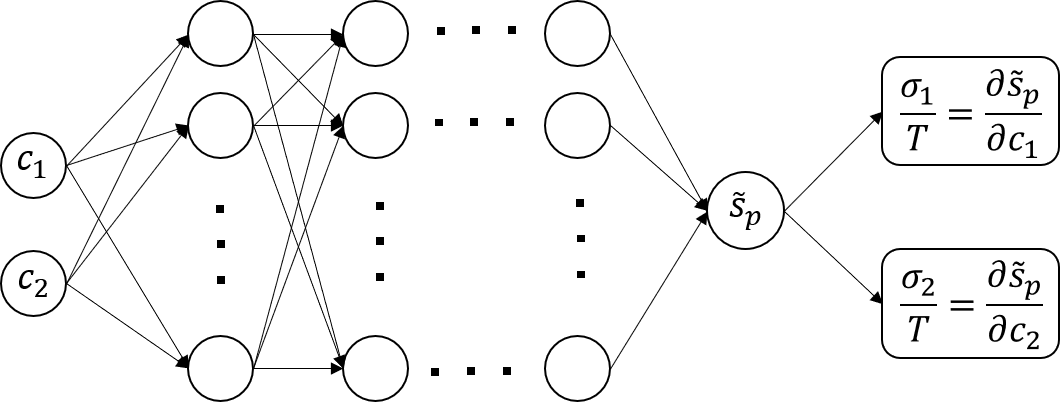}
  \caption{Sketch of the PINNs architecture.  
  }
  \label{fig:rheopinn}
\end{figure}

To demonstrate the  methodology, this paper focuses  on 2D flows,
  with the  extension to 3D flows  being straightforward. We  aim   at  representing   the  functional   form  of   the  entropy
$s_{\rm   p}(\mathsf{C}):\mathbb{R}^2\to\mathbb{R}$  as   a  function   of  the
 eigenvalues 
{$\mathsf{C}=\{c_1,c_2\}\in\mathbb{R}^2$} through a neural network of the form 
\begin{align}
  \label{NN-1}
 \tilde s_{\rm p}(\mathsf{C})= n_{\rm p} k_B \tilde s_\theta(\mathsf{C})
\end{align}
where $n_{\rm p}$ is the polymer number density and the dimensionless NN is 
\begin{align}
  \tilde s_\theta(\mathsf{C})=\left[
W_L\cdot \phi(  \cdots\phi(W_1 \cdot\phi(W_0 \mathsf{C}+b_0)+b_1 )\cdots) + b_L\right](\mathsf{C}-\mathsf{C}_{\rm eq})^2
  \label{Entropy-NN}
\end{align}
%
where
  $W_k       \in       \mathbb{R}^{h_k\times       h_{k-1}}$       and
  $b_k\in\mathbb{R}^{h_k}$ denote respectively  the weights and biases
  of the $k$-th hidden layer,  with $k\in(0,\cdots,L)$.  The number of
  nodes of the $k$-th layer is $h_k$.
The  collection of  all   trainable
parameters   of   the  network   is   identified as  
{$\theta    =    \{W_k,b_k\}_{k=0}^L$.}    The    activation    function
$\phi: \mathbb{R}  \to \mathbb{R}$  is a smooth  non-linear function
that  is  applied  element-wise to  a  multivariate argument.   The
choice of the function is  arbitrary and often problem-dependent, with
common selections for $\phi$ including the hyperbolic tangent or the
sine  function, {chosen in this work.} Finally,  the factor  involving the  equilibrium value
$\mathsf{C}_{eq}=\{1,1\}$ of  the  conformation tensor's  eigenvalues is a common approach followed when implementing PINNs in order to impose exact satisfaction of boundary and/or initial conditions \cite{arzani2023theory, sukumar2022exact}. In particular, we ensure that
$s_{\rm p}(\mathsf{C}_{\rm  eq})=0$  and ${\sigma}_\alpha(\mathsf{C}_{\rm  eq})=\{0,0\}$. 
This requirement is necessary because a non-zero entropy at the equilibrium will often result in unstable flow predictions once the neural network is coupled with RheoTool. Additional information on the expressions for $s_{\rm p}$, and therefore $\bm \sigma$, for simple models are discussed in Appendix \ref{sec:transient}.\\
A graphical description of Eq.(\ref{Entropy-NN}) is shown in Fig.  \ref{fig:rheopinn}, where it is also
indicated how the eigenvalues $\sigma_\alpha$ are obtained  from  automatic differentiation  of the NN entropy
function with respect to $\mathsf{C}$.\\

The  final  ingredient  of  a  NN  is  the  loss  function  whose
  minimization  produces  the parameters  of  the  network.  The  loss
  function  in PINNs  is  constructed in terms  of  residuals of the PDE.   
The
definition of the  residuals $e_\alpha(c_\alpha;\theta) $ follows from the \textsc{generic}-consistent
PDE (\ref{eq:refec}) that already uses the NN representation in (\ref{NN-1})
\begin{align}
 e_\alpha(c_\alpha;\theta)
 &\equiv \partial_t c_{\alpha} + v_j\partial_jc_{\alpha} -2c_{\alpha}\kappa_{\alpha\alpha} - \frac{2}{\lambda_{\rm p} } c_\alpha \frac{\partial \tilde s_\theta}{\partial c_\alpha}
\label{eq:ec-1}
\end{align}
where $\alpha= 1,2$ for 2D problems and the partial  derivative of the NN $\tilde s_\theta$ is computed
through  automatic differentiation \cite{paszke2017automatic}. The residuals in Equation \ref{eq:ec-1} can be obtained through any dataset -- produced either by simulation or experiments -- for the velocity gradient and the microstructure (i.e., the conformation tensor). 

The residuals can be further simplified when the model is trained exclusively on the steady state line.
Therefore, 
when training the PINN with synthetic viscometric data (PINN- rheometric), we will use the residual
\begin{align}
  e_\alpha(c_\alpha;\theta) = -2c_{\alpha}\kappa_{\alpha\alpha} - \frac{2}{\lambda_{\rm p}} c_\alpha \frac{\partial \tilde s_{\theta}}{\partial c_\alpha} \label{eq:ec_ss}
\end{align}

With either of the residuals defined 
by Eqns (\ref{eq:ec-1}) and (\ref{eq:ec_ss}), we define the loss function

\begin{align}
  \mathcal{L} (c_1 ^i, c_2 ^i, \theta) 
 = \frac{1}{N}\sum_{i=1}^N \Bigl( \lambda_1|| {e}_{1}(c_1 ^i;\theta) ||^2 + \lambda_2|| {e}_{2}(c_2 ^i;\theta) ||^2 \Bigr) \label{eq:loss}  
\end{align}
where $\{(c_1 ^i, c_2 ^i)\}_{i=1}^N$ represents the dataset with $N$ points used for the training of our model. The parameters $\lambda_1$ and $\lambda_2$ represent two scalars whose purpose is to balance the interplay between the two residuals. Unbalanced loss components are known to be detrimental for the training process of PINNs, which can lead to slow or unfeasible convergence \cite{wang2022whenandwhy}. We established the values of the two scalars based on the heuristics proposed in \cite{schmid2024physics}, where their value is given as the proportion of the two residuals at the first iteration. 
For the specific models we train in our study, we notice that this approach generally yields a ratio $\lambda_1/\lambda_2 \approx 10^{-4}$.  Finally,  by
minimizing the loss function in Eq. \ref{eq:loss}, we  identify the optimal NN entropy
function  that ensures  the {dynamic  equations} in (\ref{eq:ec-1}) to be consistent
with the data. We train our models by the Adam optimizer \cite{kingma2014adam}, which is an enhanced first-order stochastic optimization algorithm ubiquitous in the machine learning community.



One of the main benefit of using a PINN 
to approach the minimization of PDE residuals is the flexibility of the training formulation. Indeed, all the physical quantities included in Equation \ref{eq:ec-1} can be obtained through high-fidelty numerical simulation, 
or experimental measurements and all those data can be introduced in the same training scheme. 




\subsection{
Data-driven CFD: coupling PINN with RheoTool}

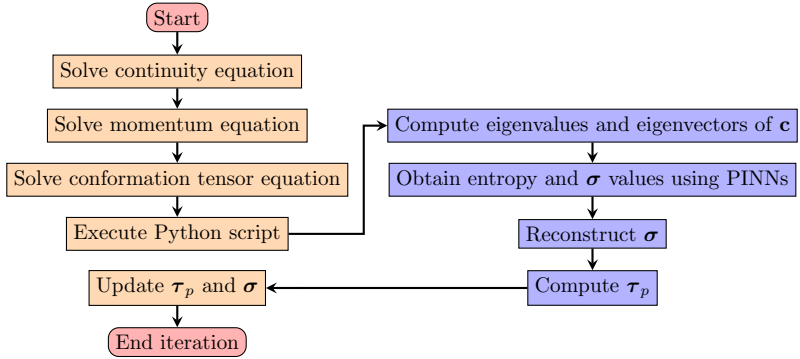
\begin{figure}[b!]
    \centering
    \begin{tikzpicture}[node distance=0.9cm, every node/.style={scale=0.8}]

    \node (start) [startstop] {Start};
    \node (step1) [process, below of=start] {Solve continuity equation};
    \node (step2) [process, below of=step1] {Solve momentum equation};
    \node (step3) [process, below of=step2] {Solve conformation tensor equation};
    \node (pythonstart) [process, below of=step3] {Execute Python script};
    \node (update) [process, below of=pythonstart] {Update $\boldsymbol{\tau}_p$ and $\boldsymbol{\sigma}$};
    \node (end) [startstop, below of=update] {End iteration};

    \node (step4) [pythonprocess, right of=pythonstart, xshift=6cm, yshift=1.8cm] {Compute eigenvalues and eigenvectors of $\mathbf{c}$};
    \node (step5) [pythonprocess, below of=step4] {Obtain entropy and $\boldsymbol{\sigma}$ values using PINNs};
    \node (step6) [pythonprocess, below of=step5] {Reconstruct $\boldsymbol{\sigma}$};
    \node (step7) [pythonprocess, below of=step6] {Compute $\boldsymbol{\tau}_p$};

    \draw [arrow] (start) -- (step1);
    \draw [arrow] (step1) -- (step2);
    \draw [arrow] (step2) -- (step3);
    \draw [arrow] (step3) -- (pythonstart);
    \draw [arrow] (update) -- (end);

    \draw [arrow] (pythonstart.east) -- ++(1.0,0) |- (step4.west);
    \draw [arrow] (step4) -- (step5);
    \draw [arrow] (step5) -- (step6);
    \draw [arrow] (step6) -- (step7);
    \draw [arrow] (step7.west) -- ++(-3.0,0) |- (update.east);

    \end{tikzpicture}
    \caption{Flowchart of the procedure for macroscopic flow simulations using RheoTool and Python script PINN integration. Orange boxes refer to RheoTool actions. Purple boxes refer to actions in python script.}
    \label{fig:flowchart}
\end{figure}

The macroscopic flow simulations in this article have been performed using the RheoTool library. RheoTool is an extension for OpenFOAM (Finite Volumes), a popular open-source computational fluid dynamics (CFD) software, designed specifically for simulating non-Newtonian flows. RheoTool was developed by Pimenta \emph{et al.} to provide advanced viscoelastic numerical methods to the OpenFOAM community \cite{PIMENTA2017}. RheoTool is publicly available and incorporates various constitutive models for non-Newtonian fluids including power-law, Carreau, Cross, Bingham, Herschel-Bulkley, and viscoelastic models like Oldroyd-B, Giesekus, and FENE-P among many others.\\
In this paper, we use a modification of the \textit{rheoFoam} uncoupled solver that incorporates a python script where the PINN model is executed using PyTorch. The RheoTool-PINN interaction is achieved using the \textit{PythonPal} interface \cite{rodriguez2022}, a header-only library that provides high-level methods for OpenFOAM (C++) to Python communication. For example, it provides ready-to-use methods such as the constructor for the python script, a \textit{passToPython} function that creates a NumPy array in the Python interpreter from OpenFOAM data, and the \textit{execute} command to run the python script.
After initiating the PyTorch library and setting required variables in the modified solver, the following steps are executed at each time iteration (see flowchart in figure \ref{fig:flowchart}). First, the continuity 
Eq. \ref{eq:eq_continuity}
and momentum  
Eq.\ref{eq:eq_momentum}
are solved using a SIMPLEC type iteration. 
%
{The term $\frac{\eta_s}{2} \left(\nabla 
\bm{v} + \nabla 
\bm{v}^T \right)$ represents the diffusive term }of the solvent stress contribution, which is solved 
implicitly. In contrast, $\nabla \cdot \boldsymbol{\tau}$ is handled explicitly. The momentum 
equation is solved using RheoTool's \textit{coupling} Both-Sides-Diffusion (BSD) stabilization technique.
After the SIMPLEC iteration on continuity and momentum equations is completed, the constitutive equation for the symmetric conformation tensor $\bm{c}$ is solved in a semi-implicit form, where the right-hand side of the equation is solved explicitly:
\begin{equation}
\label{eq_conformation_evolution}
\partial_t {\mathbf{c}} + \left( \bm{v} \cdot \nabla \mathbf{c} \right) = ( \mathbf{c} \cdot \bm{\kappa} + \bm{\kappa}^T \cdot \mathbf{c}) + \frac{2}{\lambda_{\rm p} n_{\rm p} k_{\rm B}T}\mathbf{c} \cdot \bm{\sigma}
\end{equation}
Next, a python script is executed which involves the following steps. First, the eigenvalues $\{c_1, c_2\}$ and normalized eigenvectors $\{\bm{u}_1, \bm{u}_2\}$ of the conformation $\bm{c}$ are computed in each cell and sorted by size, where $c_1 > c_2$. 
Then, $\{\sigma_1, \sigma_2\}$ values in each cell are obtained through automatic differentiation of the PINN entropy representation  computed from $\{c_1, c_2\}$. 
As a next step, 
conjugate variable $\boldsymbol{\sigma}$ is reconstructed using the eigenvalues $\{\sigma_1, \sigma_2\}$ and eigenvectors  $\{\bm{u}_1, \bm{u}_2\}$ as:

\begin{equation}
\boldsymbol{\sigma}=\sum_\alpha \sigma_\alpha \bm{u}_\alpha \bm{u}_\alpha^{T} 
\end{equation}
%
%

Finally, the non-Newtonian extra-stress $\boldsymbol{\tau}$ is computed from {Eq. (\ref{eq:tau}) as
\begin{align}
  \bm{\tau} = -\dfrac{2 \eta_p}{\lambda_{\rm p} n_{\rm p} k_{\rm B} T}(\mathbf{c} \cdot \bm \sigma)  
\end{align}
 or equivalently 

\begin{align}
  \bm{\tau} = -\dfrac{\eta_p}{\lambda}(\mathbf{c} \cdot \frac{\partial \tilde s_\theta}{\partial \bm c})  
\end{align}
}



The computed stress is then used back in the SIMPLEC iteration of RheoTool, closing the time-step loop.
A graphical sketch of the structure of the algorithm is shown in Fig.\ref{fig:flowchart}.



\section{Numerical Results}
\subsection{PINN-rheometric: training }\label{sec:PINNs_results}

\begin{figure}[ht!]
	\centering
	\includegraphics[width=8cm]{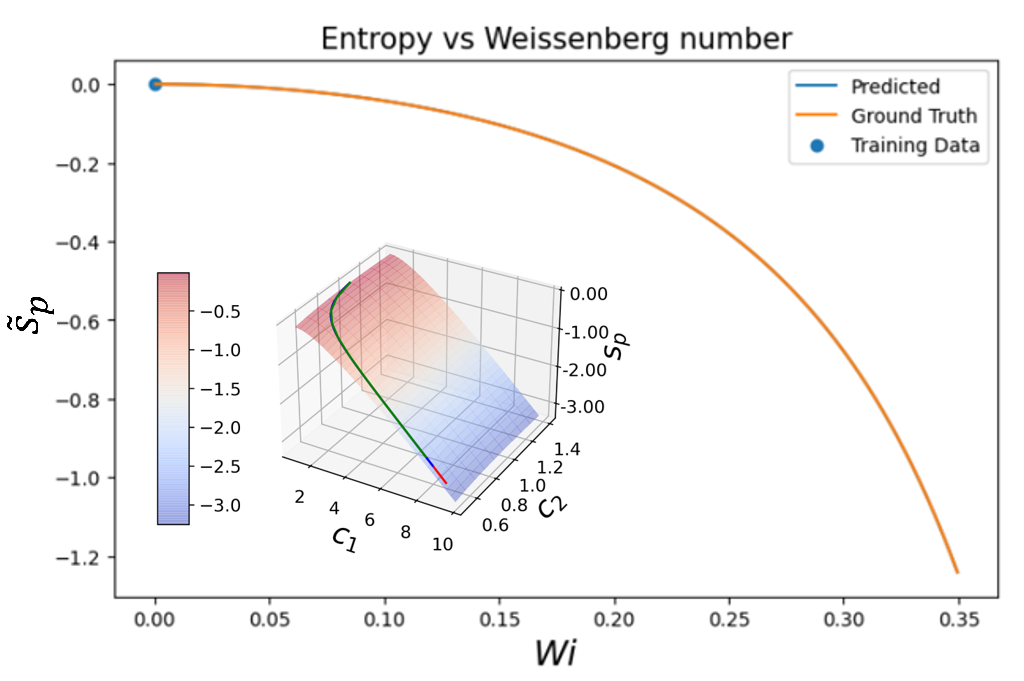}
	\caption{ Entropy as a function of Wi along the steady-state viscometric line. The inset shows the whole entropy surface (\ref{eq:SOB}) as a function of $c_1$ and $c_2$. The lines over the surface correspond to steady-state rheometric flows (extensional (red), simple shear (blue) and Poiseuille (green)). These lines all coincide.}
	\label{fig:trainingDomain}
\end{figure}

To validate the present methodology, we first use data produced by the analytical solution to Eq.(\ref{eq:const_c}) 
with (\ref{eq:eq_continuity}),
(\ref{eq:eq_momentum})
in steady-state rheometric flows. We denominate the PINN trained with  this dataset: {\em PINN-rheometric}. Since all steady-state rheometric flow solutions using the Oldroyd-B model fall on the same line in the $c_1$-$c_2$ space, we  choose to implement our training with data  for the simplest analytical solution, i.e. steady-state extensional flow
characterized by extensional rate $\dot{\epsilon}$.
For the Oldroyd-B model (OB) the analytical expression for the entropy is given by \cite{Ottinger_book}
\begin{equation}
s_{\rm p}(c_1,c_2) = \frac{k_{\rm B}}{2}(2 - c_1 - c_2 + \ln{c_1} + \ln{c_2}) \label{eq:SOB}
\end{equation}
which represents the ground truth solution to target with data, whereas in extensional flow the eigenvalues of the conformation tensor $c_\alpha$ at a given Wi are
\begin{align}
    c_{1} = & \frac{1}{1-2{\rm Wi}} \label{eq:c1_OB} \\
    c_{2} = & \frac{1}{1+2{\rm Wi}} \label{eq:c2_OB}
\end{align}
where the Weissenberg number
Wi=$\dot{\epsilon}\lambda_{\rm p}$. For  this flow, we have  computed in the Appendix \ref{sec:ss_ext}  the velocity
gradient     in    the     eigenbasis    $\kappa_{\alpha\beta}$     in (\ref{eq:kappaext}) and the residuals read,
\begin{align}\label{eq:res_explicitb}
  e^*_\alpha(c_\alpha;\theta) &= - [{\rm Wi}]_i - \frac{\partial \tilde s_\theta }{\partial c_\alpha} 
\end{align}
where the dimensionless residual $e^*_\alpha(c_\alpha;\theta)=\lambda_{\rm p} e_\alpha(c_\alpha;\theta) / 2 c_\alpha$,  $[{\rm Wi}]_i\in [0,\cdots 0.5)$  with $i=1,\cdots,N$ where $N=60,000$ is the number of data points, distributed uniformly in the interval.
Equations (\ref{eq:c1_OB}) and (\ref{eq:c2_OB}) show  that the  two eigenvalues $c_1,c_2$ can be parametrized with Wi. Furthermore, in the inset of Figure \ref{fig:trainingDomain}, we show that the data generated from  steady-state rheometric solutions of Eq. (\ref{eq:const_c})
lie on a single line in  the $c_1$-$c_2$ plane. We train our models by the Adam optimizer \cite{kingma2014adam}, which is an enhanced first-order stochastic optimization algorithm ubiquitous in the machine learning community, and we limit the training {to a maximum of} $3\cdot10^5$ iterations.

Figure \ref{fig:trainingDomain}  shows the prediction of the entropy  over the training domain ({steady viscometric} line). As expected, the PINN model is in {very good} agreement with the ground truth OB model over the training domain. Predictions for the entropy and $\bm \sigma$ outside the training domain are compared to the OB model in Section \ref{sec:PINNs_results}. 
The inset to Figure  \ref{fig:trainingDomain} shows the entropy surface $s(c_1,c_2)$ in the $(c_1,c_2)$ domain. 
The surface displays the paths explored by rheometric flows, with extensional flows highlighted in red, simple shear in blue, and Poiseuille flows in green. They all fall on top of each other.

\begin{figure}[ht!]
	\centering
	\includegraphics[width=1.0 \textwidth]{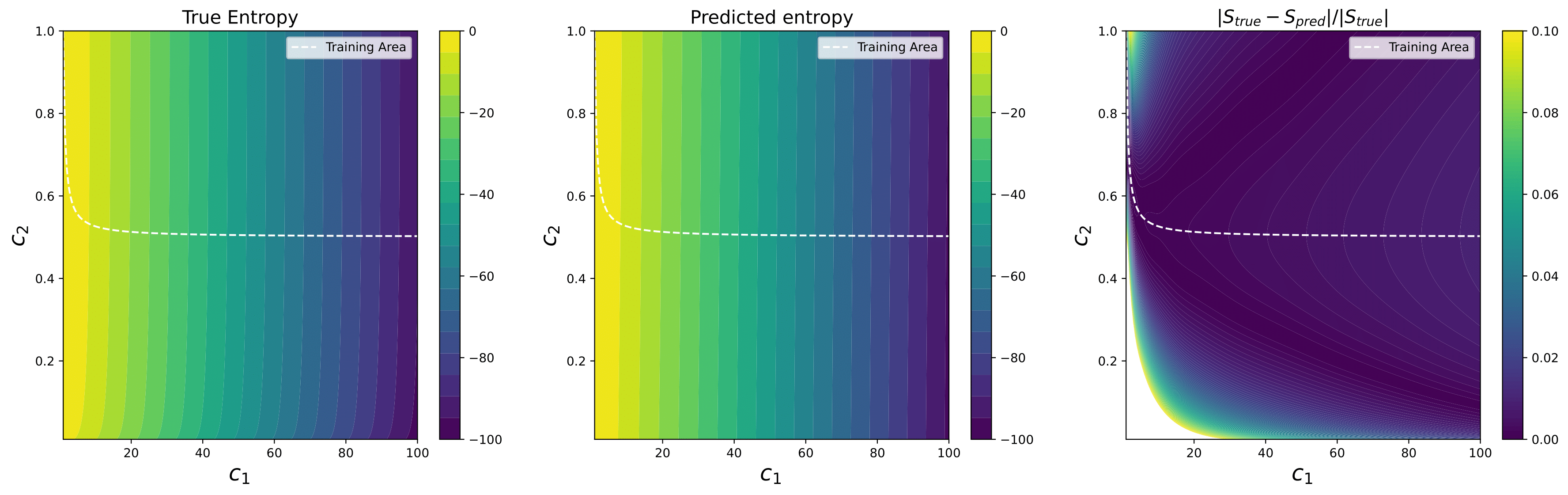}
	\caption{{\bf Left:} True entropy.
  {\bf Center:} Predicted entropy given by OB model in Eq. (\ref{eq:SOB}). {\bf Right:} 
  Relative error in the prediction of the entropy for the PINN-rheometric model.
  }
	\label{fig:s_pred}
\end{figure}

Figure \ref{fig:s_pred} shows a comparison of the PINNs prediction and the ground-truth Oldroyd-B model over the entire $(c_1,c_2)$ plane. As already shown in Figure \ref{fig:trainingDomain}, predictions are excellent over the steady-state flow line that is used for the training. The quality of the prediction is reduced as the distance to the training line is increased. 
It is important to note that some of the areas  in this graph might not be physically relevant. For example, we cannot increase $c_1$ keeping $c_2=1$ and vice-versa. Typically, when a flow becomes more `transient' or more `complex' the line representing steady-state rheometric flows (Figure \ref{fig:trainingDomain}) will start to widen leading to  an area around said line. Notably, transient extensional and shear flows only explore the region below the steady-state rheometric line (See Appendix \ref{sec:transient}), while complex flows explore both  regions above and below.   \\

\begin{figure}[ht!]
	\centering
	\includegraphics[width=1.0 \textwidth]{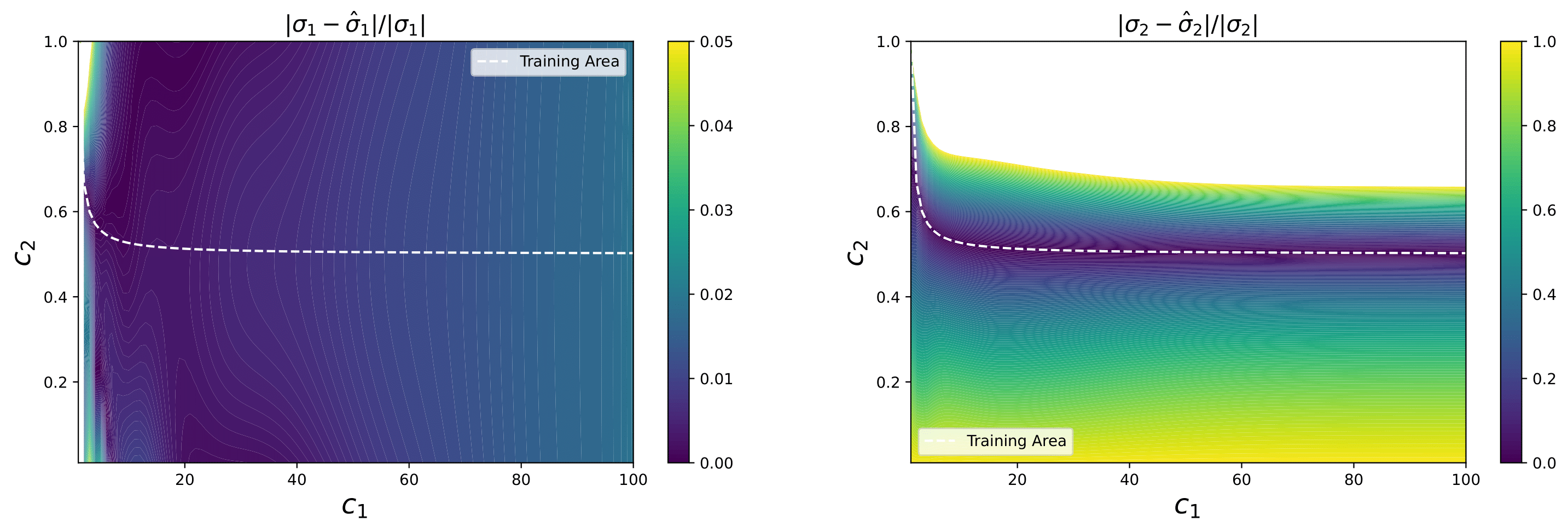}
	\caption{{\bf Left:} Relative error in the prediction of the first eigenvalue of $\bm \sigma$. {\bf Right:} Relative error in the prediction of the second eigenvalue of $\bm \sigma$. The eigenvalues of $\bm \sigma$ are determined through automatic differentiation of the PINN-rheometric model entropy in Figure \ref{fig:s_pred}.}
	\label{fig:sigma_pred}
\end{figure}

Finally, Figure \ref{fig:sigma_pred} shows the relative error in the predictions of the eigenvalues of $\bm \sigma$. The relative error in $\sigma_1$ is in general much lower than in $\sigma_2$. This is a result of the low curvature of the entropy surface and the different ranges covered by $c_1 = [1,100]$ and $c_2=[1,0.5]$ during training. This significant difference was limited during the training using augmentation in the computation of the loss function as described in Section \ref{sec:loss}. 
Nevertheless, we can observe that, while the relative error in $\sigma_1$ is kept below 5\% over the entire studied domain, the error in $\sigma_2$ increases over 50\% at a relatively short distance from the training line. This will have important consequences in the simulation of complex flows that explore wider regions of the $c_1$-$c_2$ space.

%
%

\subsection{PINN-rheometric: flow around a cylinder}

The validation of the RheoTool software for a flow around cylinder has already been reported by Alves \emph{et al.} \cite{Alves2001} in a detailed description of the computational setup required to simulate Upper Convected Maxwell (UCM) and OB fluids. In this work the fluid characteristic values to replicate the cited case have been used, that is: $\rho=1$, $\eta_s=0.59$, $\eta_p=0.41$.
\begin{figure}[ht!]
	\centering
    \begin{subfigure}[b]{0.7\textwidth}
        \includegraphics[width=0.8\textwidth]{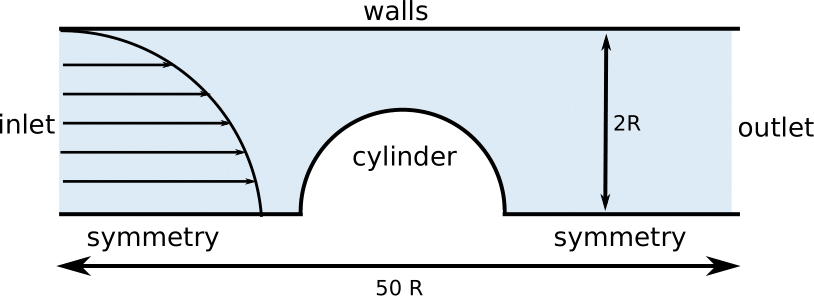}
        \caption{}
	      \label{fig:Cyl_sketch}
    \end{subfigure}
  \hfill
    \begin{subfigure}[b]{0.6\textwidth}
        \includegraphics[width=0.8\textwidth]{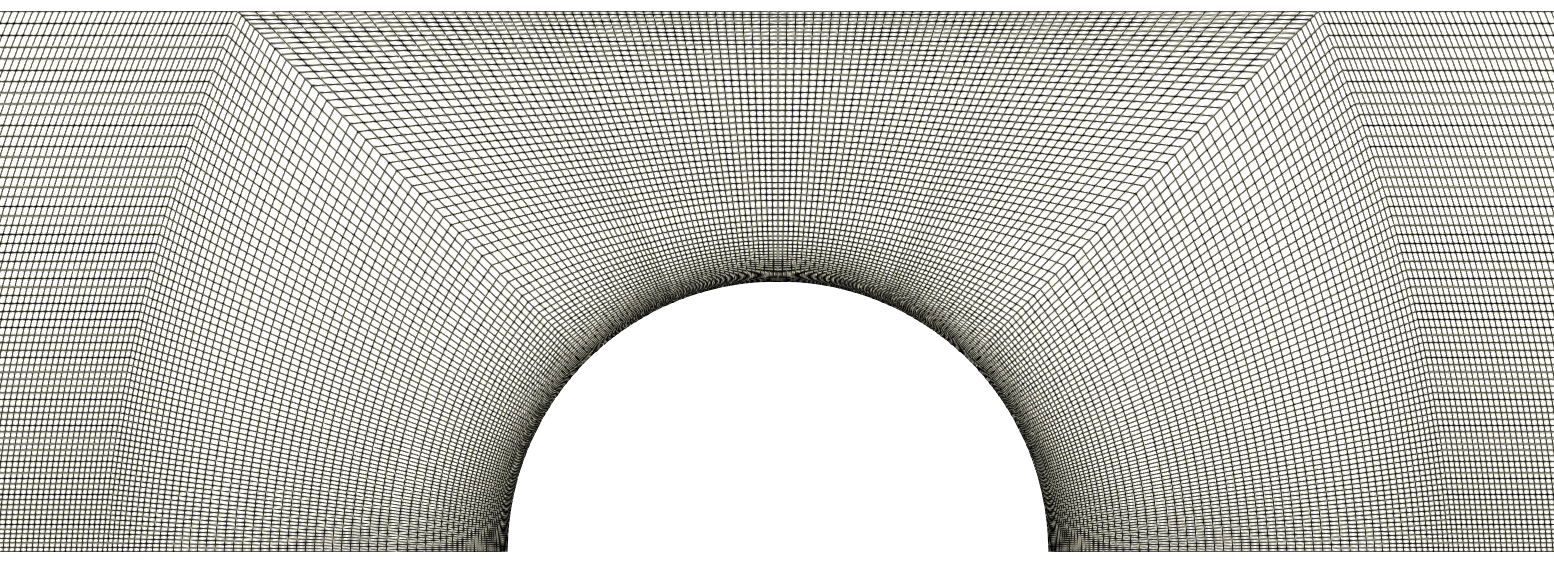}
        \caption{}
	      \label{fig:Cyl_mesh}
    \end{subfigure}
	  \caption{Sketch (a) and mesh example (b) near the cylinder in the flow around cylinder case}
   	\label{fig:Cyl_figure}
\end{figure}
A sketch of the flow geometry is presented in Figure \ref{fig:Cyl_sketch}. Full domain length is 50
times the cylinder radius, and only half of the domain is discretized as described in 
Ref.  \cite{Alves2001}. 
The
channel height is discretized with 70 cells at an expansion ratio of 3.
The employed structured mesh contains 119,422 points following the same discretization strategy as in their article.
 Fig.\ref{fig:Cyl_mesh} shows the mesh employed. 

Fluid enters from the left patch with a parabolic profile and maximum velocity $U=1$.
On the walls, no-slip conditions are used for velocity, linear extrapolation for the conformation tensor and
zero-gradient for pressure. Fluid exit is achieved by setting a zero pressure value on the right patch
and zero-gradient for the rest of the variables.
The Weissenberg number is defined consistently with the work in Ref.  \cite{Alves2001} 
using the average inlet velocity (Wi=$\overline{U} \lambda/R$). 
 In order to validate the model, simulations are run leading to a {cylinder} drag coefficient
$C_d=118.94$ obtained at Wi=0.5, consistent with the original work ($C_d=118.838$) \cite{Alves2001,Claus2013}, {where $C_d$ is calculated as:}
\begin{equation}
C_d \equiv \dfrac{1}{\eta \overline{U}}\int_s \left( \bm{\tau}_{\rm tot} - p \mathbf{I}\right)  \cdot \mathbf{n} \cdot \hat{\imath} dS
\end{equation}
{Here $\bm{\tau}_{\rm tot}$ is the sum of both the non-Newtonian and the Newtonian viscous contributions to the stress, $\mathbf{I}$ is the identity tensor, $\mathbf{n}$ is the outward cylinder suface unit normal vector, $p$ is the pressure and $\hat{\imath}$ is the unit vector in the $x$-direction.} 
{In the rest of this work, the magnitudes of tensors are evaluated using the Frobenius norm, that is 
$\tau = \sqrt{\bm{\tau} : \bm{\tau}}$,
whereas relative errors are computed  
by normalizing them by the Frobenius norm of the finite volume RheoTool reference solution. To prevent the relative error from 
diverging
when the reference solution approaches zero, a threshold value is added in the denominator. Specifically, the error in the stress variable is computed as follows:

\[
\tau_{\text{rel error}} = \dfrac{\sqrt{\bm{\tau}_{\text{abs error}} : \bm{\tau}_{\text{abs error}}}}{\tau^2 + 10^{-2}}
\]
}

\begin{figure}[hb!]
	\centering
 	\includegraphics[width=1.0 \textwidth]{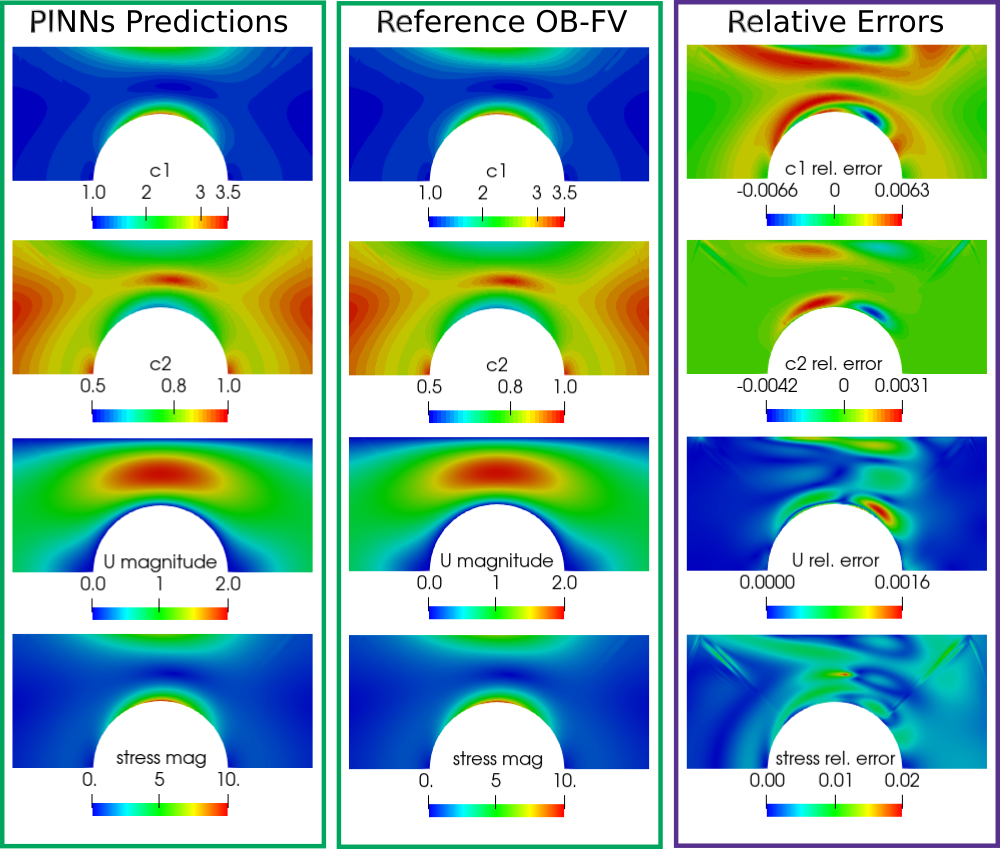}
	\caption{{RheoTool simulation with steady-state rheometric training (PINN-rheometric) vs standard RheoTool simulation of the OB model at Wi$=0.15$.}}
	\label{fig:FAC0p1}
\end{figure}
\begin{figure}[ht!]
	\centering
	\includegraphics[width=1.0 \textwidth]{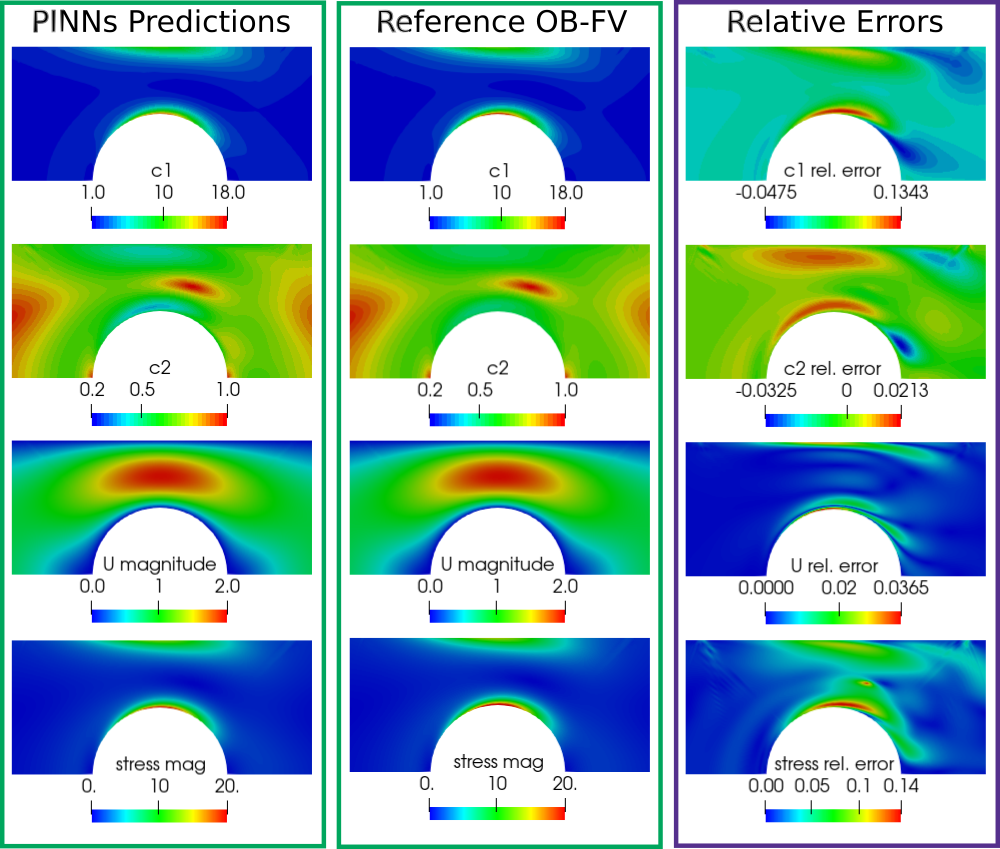}
	\caption{{RheoTool simulation with steady-state rheometric training (PINN-rheometric) vs standard RheoTool simulation of the OB model at Wi$=0.45$.}}
	\label{fig:FAC0p3}
\end{figure}
\begin{figure}[!tbp]
  \begin{subfigure}[b]{0.5\textwidth}
    \includegraphics[width=\textwidth]{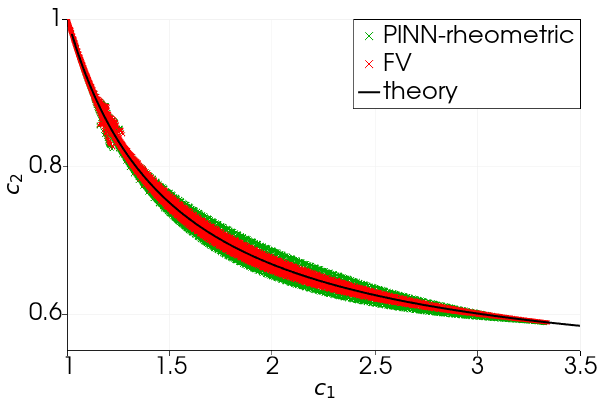}    
    \caption{Wi$=0.15$ }
    \label{fig:f1}
  \end{subfigure}
  \hfill
  \begin{subfigure}[b]{0.5\textwidth}
    \includegraphics[width=\textwidth]{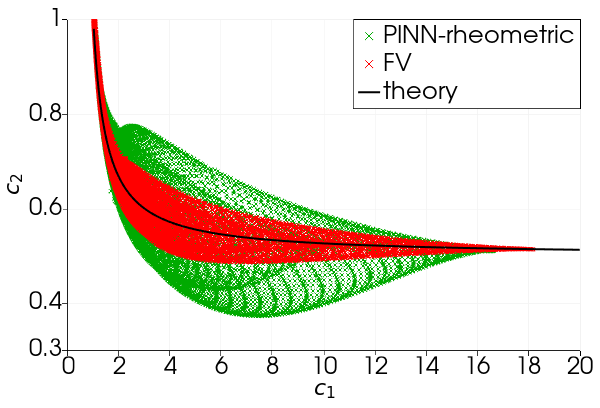}
    \caption{Wi$=0.45$}
    \label{fig:f2}
  \end{subfigure}
  \caption{$c_1$-$c_2$ region covered in simulations of flow around a cylinder with Wi$=0.15$ (a) and Wi$=0.45$ (b). Green crosses represent the results using the PINN-rheometric model, red crosses represent the Oldroyd-B implementation in RheoTool (FV), and the black solid line represent the analytical solution for steady-state rheometric flows where the training has been applied.}
  \label{fig:fig10}
\end{figure}

Figures \ref{fig:FAC0p1} and \ref{fig:FAC0p3} show the simulation results for flow around a cylinder comparing predictions of the {PINN-rheometric} model and the ground-truth OB model. For this comparison, we monitor the fields $c_1$, $c_2$, the magnitude of the velocity and stress fields in the simulation domain. PINN-rheometric model flow predictions errors are below 2\% for the stress and $<1\%$ for all other fields at low Wi$=0.15$ as shown by Figure \ref{fig:FAC0p1}. However, as Wi is increased to Wi$=0.45$ (See Figure \ref{fig:FAC0p3}) significant relative errors exceeding 10\% for the stress and $c_1$ are observed. This is a result of the large error in $\sigma_2$ reported in Fig.\ref{fig:sigma_pred}. 
The errors in the stress and  $\sigma_\alpha$ are linked to the spreading of the region of the conformation space explored during simulation (See Fig.\ref{fig:f2}).\\
These results can be better appreciated in Figures \ref{fig:f1} and \ref{fig:f2}. For low Wi, the conformation tensor eigenvalues are very close to the steady-state line. As Wi increases the explored region in the $c_1$-$c_2$ plane becomes wider and separates from the steady-state line. The error  in $\bm \sigma$ is larger in this region where the PINN model is extrapolating away from the steady viscometric training domain. As a result, the {poor} estimation of the stress in the simulation {numerically} propagates these errors, which in turn contribute to create an even wider region of the {mapped conformational} space (i.e., the maximum for $c_1$ is higher and the minimum for $c_2$ is lower {than the ground truth OB solution}).
%

%
Solutions for the stress on the symmetry plane and around the cylinder are also reported following the benchmark studied in \cite{Alves2001}.
It can be observed in Figure \ref{fig:fig_PINN_stress} that at high Wi number the stress on top of the cylinder is increased due to high error
{associated to large} $c_1$ values. From these results, it can be {concluded} that {in order} to investigate regions outside the line in the $c_1-c_2$ space explored in rheometric steady-state flow, 
one should train the network with data from
non-rheometric flows (i.e., either complex or transient flows cases able to map a wide range of the conformation space). Examples of the conformational space explored in complex flow around a cylinder are presented in Fig. \ref{fig:fig10}, whereas the cases corresponding to transient extensional and shear flows are presented in Appendix \ref{sec:transient}.

\subsection{PINN-complex: training}

In this section, we study a PINN constitutive model trained with data from macroscopic simulations. This allows us to include in the training values outside the line explored by rheometric steady-state flows. To that end, the data obtained for the steady-state flow around cylinder at Wi$=0.45$ has been processed to evaluate the residuals in Eq. (\ref{eq:ec-1}). Effectively, we mimic the procedure that would be followed in the application of our learning protocol to experimental data (i.e., with the conformation tensor and velocity field measured via PIV \cite{Haward2021}). Here the PINNs model is trained exclusively using the discrete in-silico data obtained from CFD RheoTool simulation of the OB model,  but notably, the PINN is agnostic of the ``true" constitutive model {behind the training data}.

{The} procedure starts by {computing the eigenvalues and eigenvectors}  from the conformation tensor.
{As we work with steady-state flow the} time derivative is zero for steady-state converged solution, $\kappa_{\alpha
\alpha}$ is computed using definition (\ref{e4}).
The {required} gradients of these vectors are computed using linear Gauss method. With all terms evaluated, the derivative of the entropy with respect to the conformation tensor is computed by Eq. (\ref{eq:ec-1}). Effectively this is equivalent to computing $\sigma_{\alpha}$, which is then used for the PINN-complex training with PyTorch. 

\begin{figure}[ht!]
	\centering
	\includegraphics[width=1.0 \textwidth]{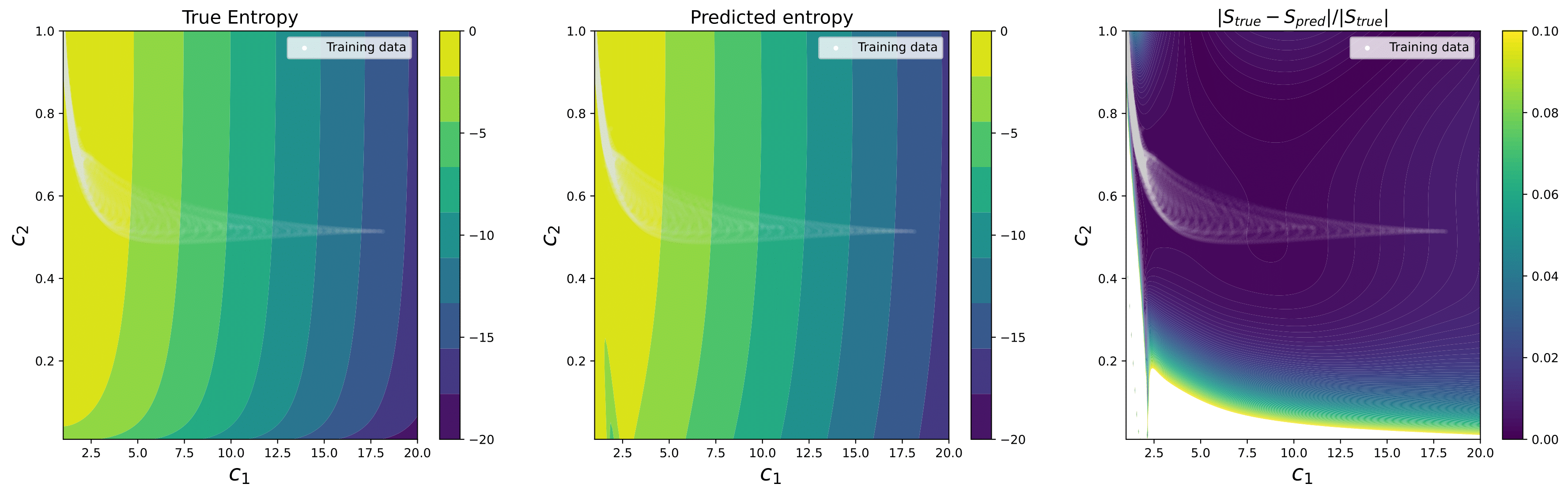}
	\caption{{\bf Left}: True entropy given by OB model in Eq. (\ref{eq:SOB}).
 {\bf Center:} Predicted entropy. {\bf Right:} Relative error in the prediction of the entropy for the PINN-complex model.}
	\label{fig:s_pred_data}
\end{figure}

In Fig. \ref{fig:s_pred_data} we compare the {entropy} predictions of the PINN-complex model against the ground-truth solution {given by} the Oldroyd-B model. {As in the PINN-rheometric model, the predictions for the entropy given by PINN-complex} are very accurate 
in the training area of the conformational space and a relative error is consistently below $10\%$ 
{even in regions far from the training area.}\\

\begin{figure}[ht!]
	\centering
	\includegraphics[width=1.0 \textwidth]{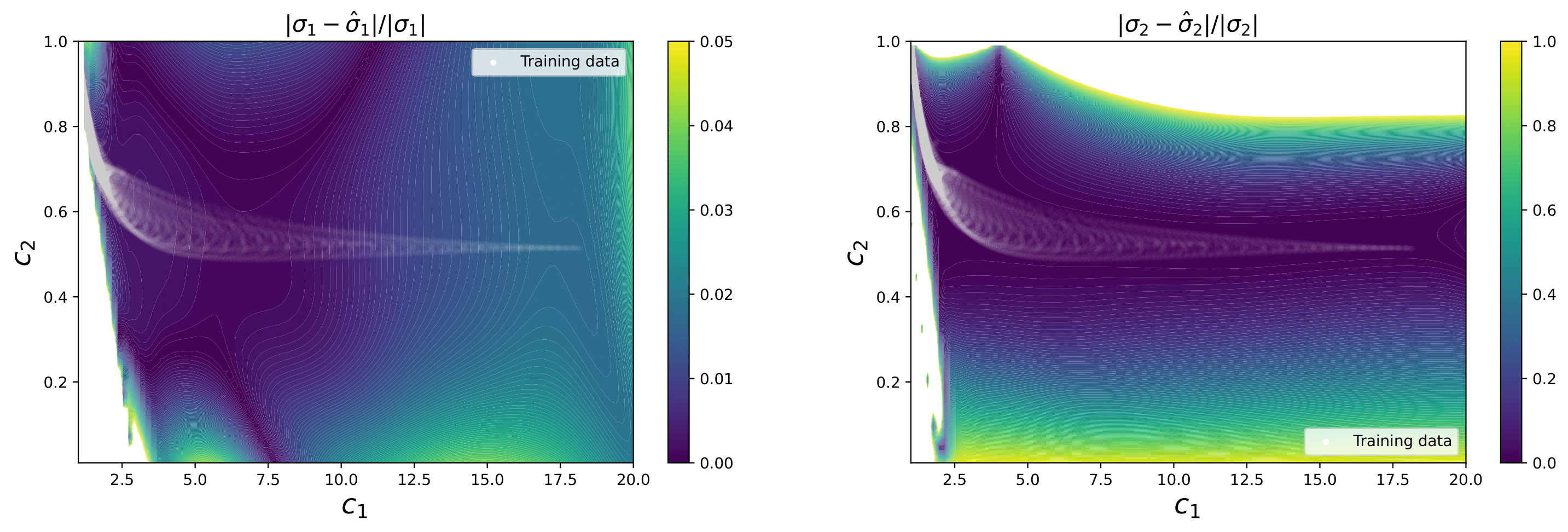}
	\caption{{\bf Left:} Relative error in the prediction of the first eigenvalue of $\bm \sigma$. {\bf Right:} Relative error in the prediction of the second eigenvalue of $\bm \sigma$. The eigenvalues of $\bm \sigma$ are determined through automatic differentiation of the PINN-complex model entropy in Fig. \ref{fig:s_pred}. 
 }
	\label{fig:sigma_pred_data}
\end{figure}

Fig. \ref{fig:sigma_pred_data} shows the relative error achieved by the PINN-complex model for the prediction of the eigenvalues of $\bm \sigma$. The relative errors in $\sigma_1$ and $\sigma_2$ are again affected by the low curvature of the entropy surface due to their different training ranges $c_1 = [1,20]$ and $c_2=[1,0.5]$. However, we notice that introducing training data from complex flows can largely benefit the accuracy of prediction and goodness of extrapolation for the value of $\sigma_2$.

\subsection{PINN-complex: {flow around a cylinder}}

{In this section we compare the accuracy of the RheoTool {flow simulations using the  PINN-complex model discussed in the previous section} against the 
PINN-rheometric model and the ground truth OB solution {(i.e., the Oldroyd-B model used to generate the training datasets)}.
We consider here the most challenging case of Wi=0.45 where significant errors in the stress and conformation tensor fields were 
observed {when using the PINN-rheometric model.}
}
%
\begin{figure}[ht!]
	\centering
	\includegraphics[width=1.0\textwidth]{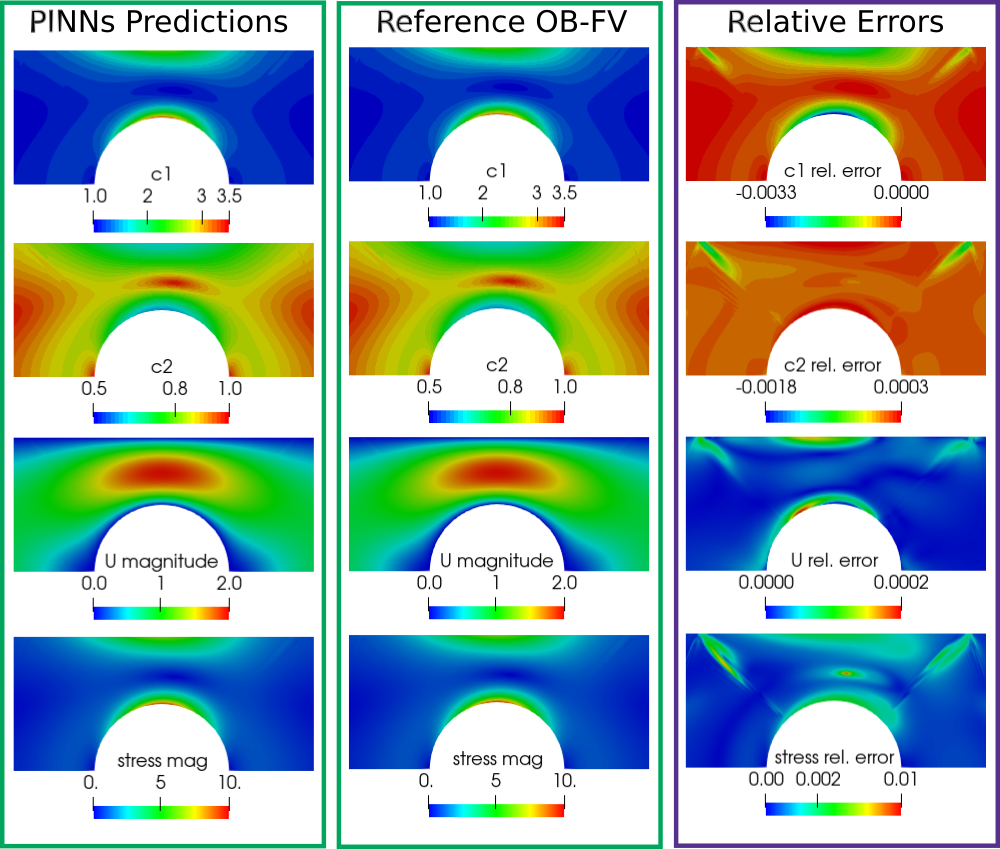}
	\caption{Comparison of RheoTool simulation results for PINN-complex and Oldroyd-B models at Wi$=0.15$. 
 }
	\label{fig:complex_solution_wi0p1_and_wi0p15}
\end{figure}
\begin{figure}[ht!]
	\centering
	\includegraphics[width=1.0\textwidth]{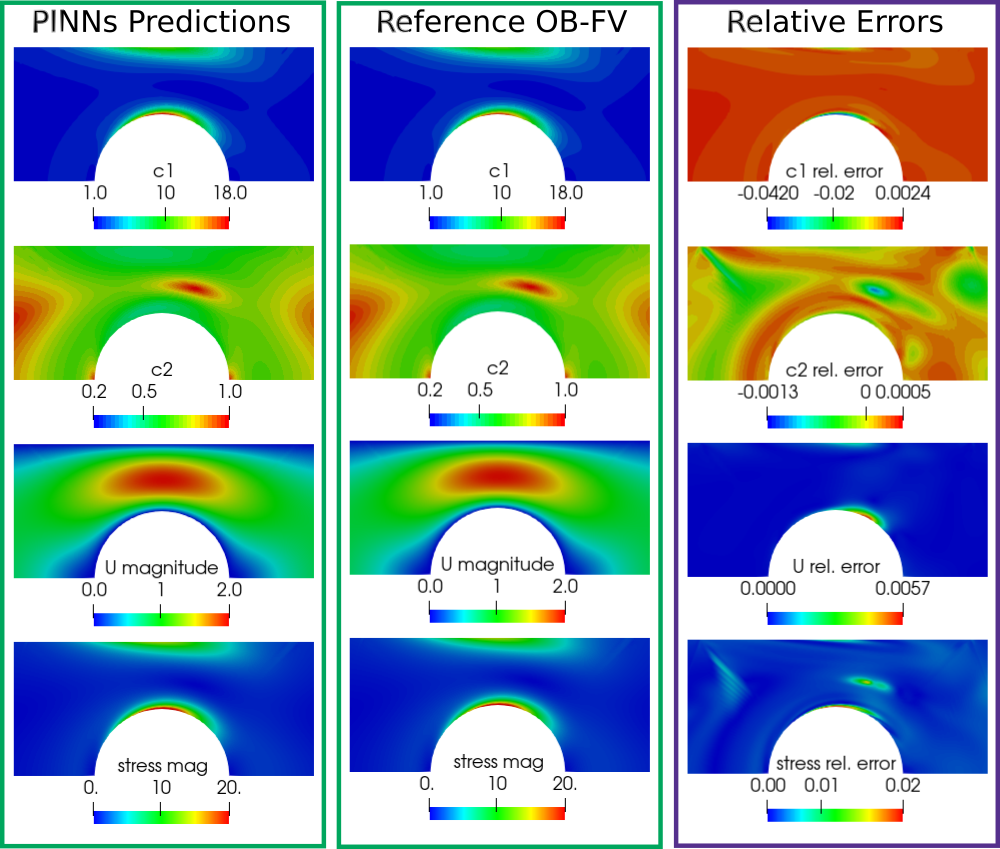}
	\caption{
Comparison of RheoTool simulation results for PINN-complex and Oldroyd-B models at Wi$=0.45$.
 }
	\label{fig:complex_solution_wi0p3_and_wi0p45}
\end{figure}

{The new RheoTool results {using the PINN-complex model} are reported in Figures \ref{fig:complex_solution_wi0p1_and_wi0p15} and \ref{fig:complex_solution_wi0p3_and_wi0p45}.
Comparing these results with the previous ones in Fig. \ref{fig:FAC0p3}, 
we can clearly observe a significant overall improvement in the accuracy {in the predictions for the eigenvalues of the conformation tensor and the stress}.
In particular, relative error in $c_1$ shows the maximum error in the complex case located on top of the cylinder (where $c_1$ is also maximum) with a value of 4.2\%, whereas for the PINN-rheometric in Fig. \ref{fig:FAC0p3} was around 13.4\%. The error in the rest of the domain for the PINN-complex 
 is generally below 1\%, far better than the 
PINN-rheometric simulation {that has an error exceeding 5 \%}.
Also the PINN-complex 
simulation results for $c_2$ are significantly better, 
 with relative error under 0.1 \%. On the contrary, the  PINN-rheometric case in Fig. \ref{fig:FAC0p3} shows error as large as 3 \%.}

\begin{figure}[!tbp]
\centering
    \includegraphics[width=0.7\textwidth]{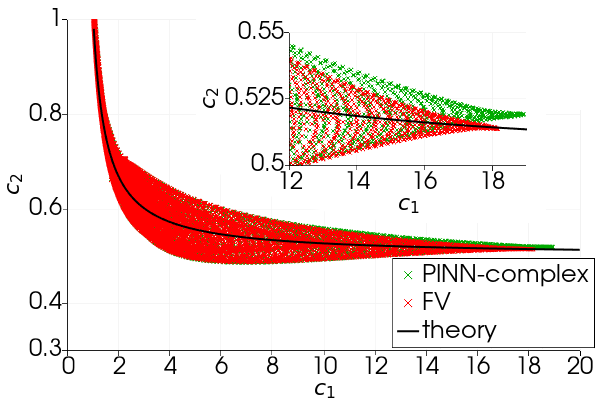}
    \caption{$c_1$-$c_2$ region covered in simulation of flow around a cylinder with Wi$=0.45$. Green crosses represent the results using the {PINN-complex model}, red crosses represent the OB implementation in RheoTool (FV), whereas the black solid line represent the analytical solution for steady-state rheometric flow.}
    \label{fig:f3}
\end{figure}

The reasons behind the improvement of the constitutive modeling using complex flow data can be further analyzed comparing Figures \ref{fig:f2} and \ref{fig:f3}. Here the regions covered in the $c_1-c_2$ space using the different training datasets are compared. Fig. \ref{fig:f2} reveals the widening of the $c_1-c_2$ space of the {PINN-rheometric} predictions outside of the area explored by the {ground truth RheoTool solution for the  Oldroyd-B model}. This result reveals how the larger error in the stress has a strong effect on the conformational space {that is explored in a given simulation}. In contrast, no widening of the explored area in the $c_1-c_2$ space is observed in Fig. \ref{fig:f3} {(i.e., PINN-complex simulation results for $c_1$ and $c_2$ throughout the simulation domain are in very good agreement with the Oldroyd-B model simulation results).} The maximum error occurs at the top of the cylinder, where $c_1$ reaches its peak, approaching the maximum $c_1$ value used for the training data. This result {clearly confirms} that the training with the complex flow data 
better captures the underlying constitutive model (Oldroyd-B) {in terms of consistent conformation space mapping for a complex flow problem.}\\

\begin{figure}[ht!]
         \centering
         \includegraphics[width=\textwidth]{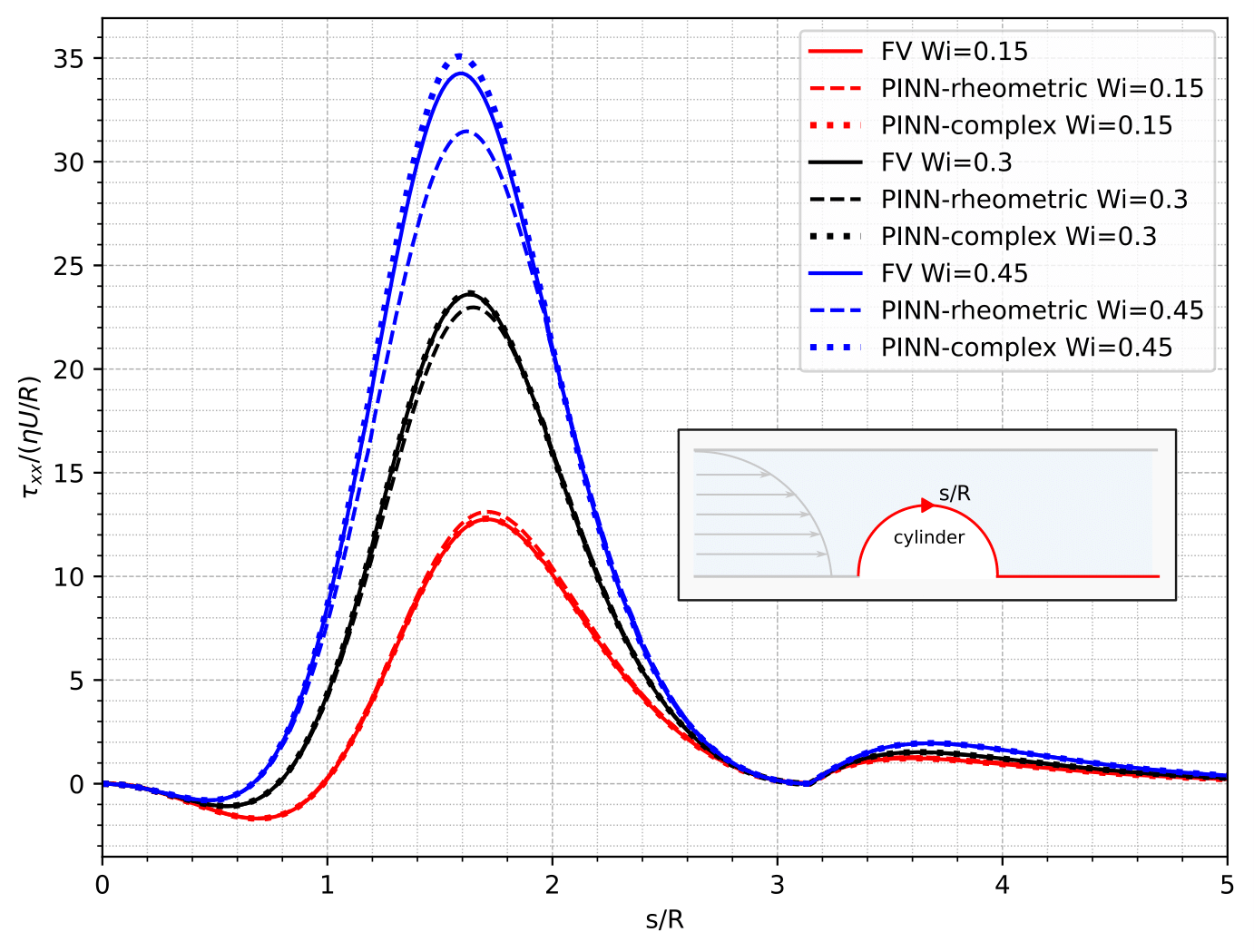}
         \caption{Comparison of stress on cylinder and symmetry plane. Inset shows the line over which the stress is computed.}       
         \label{fig:fig_PINN_stress}
\end{figure}

{Finally, 
in Fig. \ref{fig:fig_PINN_stress}
we examine the stress values along the channel mid-plane and on the cylinder surface.} The figure presents numerical results 
using three approaches: {PINN-complex, PINN-rheometric, and the reference finite volume Oldroyd-B solution computed with the standard RheoTool.} At Wi=0.45, the PINN-complex model shows 
significantly {improved} results compared to the {PINN-rheometric} model. Furthermore, {we also observe that} by reducing the Weissenberg number to 0.3, an excellent agreement is achieved in the PINN-complex case. This improvement is due to the lower maximum $c_1$ simulated, hence avoiding the $c_1$ limit used during training. {As expected, for Wi$=0.15$ both models, PINN-rheometric and PINN-complex, provide a good representation of the solution computed with Oldroyd-B model.}

\clearpage

\subsection{PINN-complex: flow around an array of cylinders}

\begin{figure}[ht!]
	\centering
    \begin{subfigure}[b]{0.4\textwidth}
         \centering
         \includegraphics[width= \textwidth]{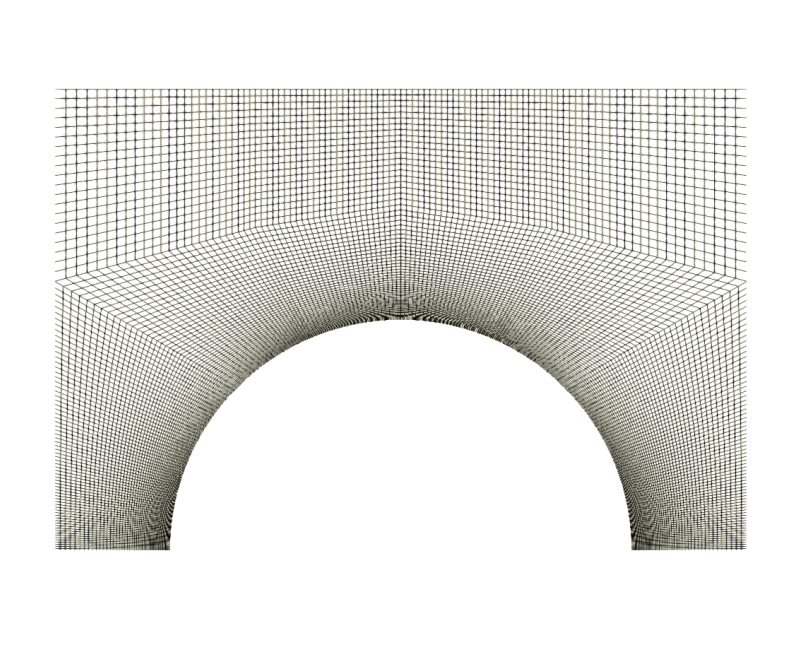}
         \caption{PAC mesh for L=3R}
         \label{fig:cyclic_mesh}
    \end{subfigure}
    \begin{subfigure}[b]{0.57\textwidth}
         \centering
         \includegraphics[width= \textwidth]{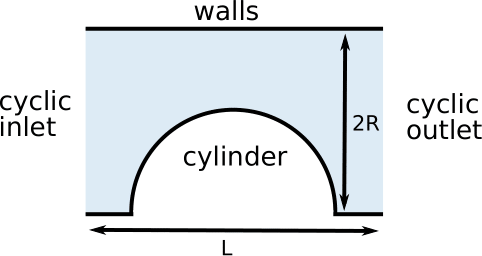}
         \caption{PAC sketch}
         \label{fig:cyclic_sketch}
    \end{subfigure}
         \caption{Periodic array of cylinder (PAC) test case geometry representation.}
         \label{fig:cyclic_geometry}
\end{figure}

\begin{figure}[b!]
	\centering
     \begin{subfigure}[b]{0.48\textwidth}
         \centering
         \includegraphics[width=\textwidth]{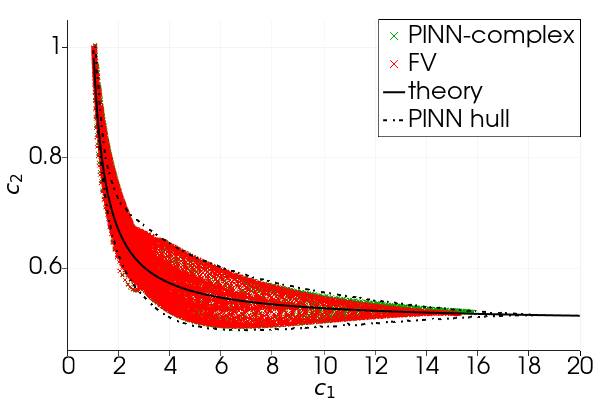}
         \caption{L=2.5 R}
         \label{fig:steady_solution_wi0p4_hull_L2p5}
     \end{subfigure}
      \begin{subfigure}[b]{0.48\textwidth}
         \centering
         \includegraphics[width=\textwidth]{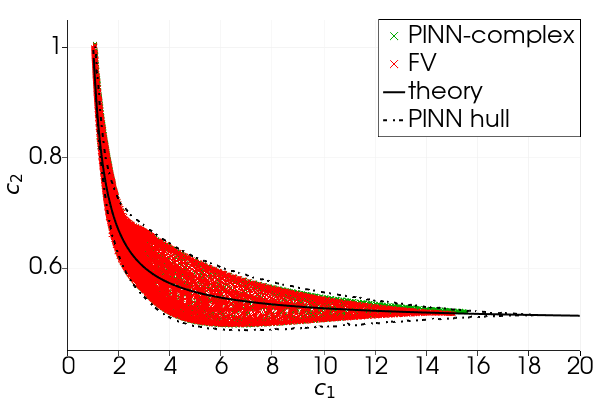}
         \caption{L=3 R}
         \label{fig:steady_solution_wi0p4_hull_L3}
     \end{subfigure}
         \caption{$c_1$-$c_2$ region covered in a PAC flow at Wi$=0.4$ at (a) L=2.5 R and (b) L=3 R for the PINN-complex model (green crosses). Red crosses represent the Oldroyd-B implementation in RheoTool (FV), and the black solid line represent the analytical solution for steady-state rheometric flows. The convex hull of the PINN-complex training range in a flow around {a single} cylinder is also shown with a dashed line.}
         \label{fig:steady_solution_wi0p4_hull_L}
\end{figure}

One important advantage of the data-driven procedure {to model the constitutive equation for the stress using PINNs} presented in this work is that 
the resulting PINN models are geometry-agnostic. This means that generalization to new geometries is in principle straightforward since it does not require extra training. 
In order to assess this feature of the model,
in this section we applied the 
PINN-complex model to a new geometry
setup: the flow around a periodic array of cylinders (PAC)\cite{Ellero2011, Nieto2022}. {In this case a single cylinder is simulated within the unit cell and periodic boundary conditions on the left/right faces are activated,
making therefore possible for the cylinders to interact hydrodynamically. This will be important, since for significantly close cylinders, a topological change of the flow occurs, therefore enabling the assessment of the PINN accuracy in a different flow scenario.}  Two different separation lengths between cylinders $L$ are used for our analysis: $L=3R$ and $L=2.5R$. {In the latter case a topological change with a flow separation instability in the space between cylinders occurs.} Fig. \ref{fig:cyclic_geometry} shows the geometry and mesh details.
In Fig. \ref{fig:steady_solution_wi0p4_hull_L2p5} the region explored by the simulation results of the PAC flow for PINN-complex and RheoTool of 
Oldroyd-B model (FV) are shown for Wi=0.4. At this Wi, the explored area in the $c_1-c_2$ space for the new PAC geometry is within the training region (PINN hull) that is represented by the dashed line in Fig. \ref{fig:steady_solution_wi0p4_hull_L2p5}.

\begin{figure}[ht!]
	\centering
	\includegraphics[width=0.9 \textwidth]{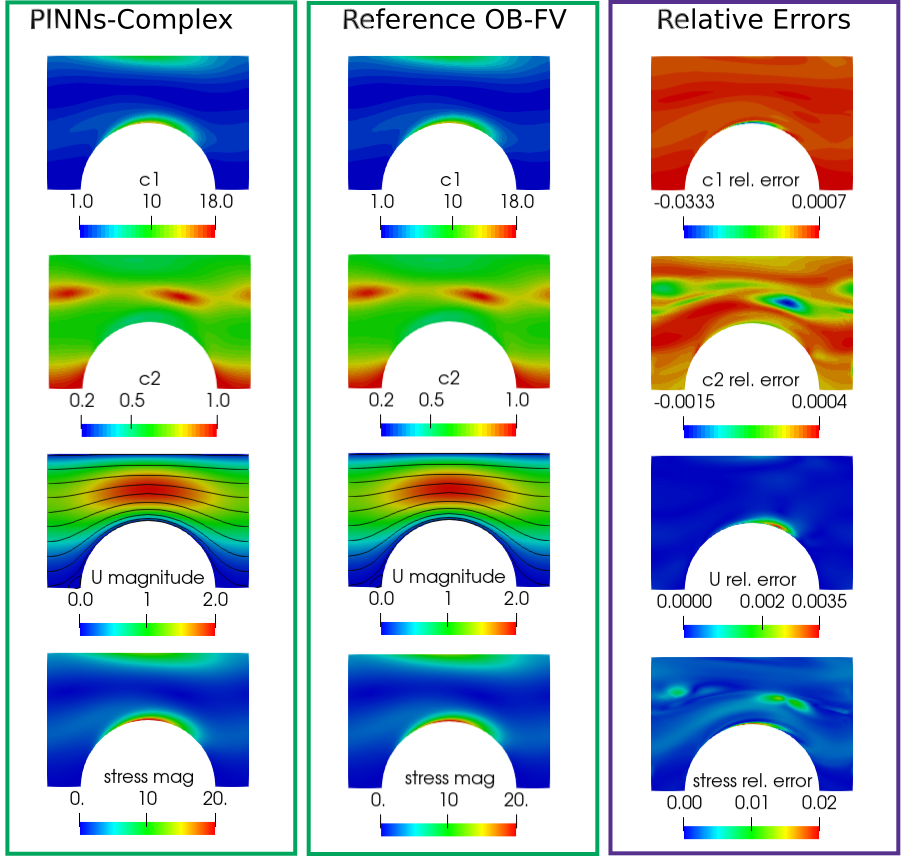}
	\caption{{Comparison of}
 RheoTool simulation with PINN-complex {and} 
 Oldroyd-B models at Wi$=0.4$ in a periodic array of cylinders (PAC) for $L=3R$.
 }
	\label{fig:train_solution_wi0p4}
\end{figure}

Figure \ref{fig:train_solution_wi0p4} shows a comparison of the different field $(c_1, c_2, U, \tau)$ in simulations for the PAC flow with $L=3R$. The PINN-complex predictions demonstrate good accuracy, as indicated by the relative errors in $c_1$ and $c_2$ remaining below 3\% and 0.15\%, respectively. The relative errors in stress and velocity are also small. However, it is important to note that the relative error for $U$ 
{is extremely sensitive near stagnation lines.}

{The same analysis is performed in the case of closely interacting cylinders for $L=2.5R$.}
In Fig.\ref{fig:train_solution_wi0p4_L25}, we analyse the relative error with the PINN-complex 
applied to this new flow scenario.
In this situation, a topological change of the flow occurs {due to the re-circulation area 
generated in the region between cylinders, as observed by the new streamlines. 
} 

{
The relative errors found in this flow type are of the same order of magnitude as for the $L=3R$ case and follow  {a similar} distribution over the fields. 
The only noteworthy difference is the increase in relative error up to 0.3\% in the $c_2$ field 
within the re-circulation area, adjacent to the cyclic inlet/outlet. 
{Note that}, the plot ranges for $c_2$ are kept the same as the $L=3R$ case to facilitate a direct comparison.

\begin{figure}[ht!]
	\centering
	\includegraphics[width=0.9 \textwidth]{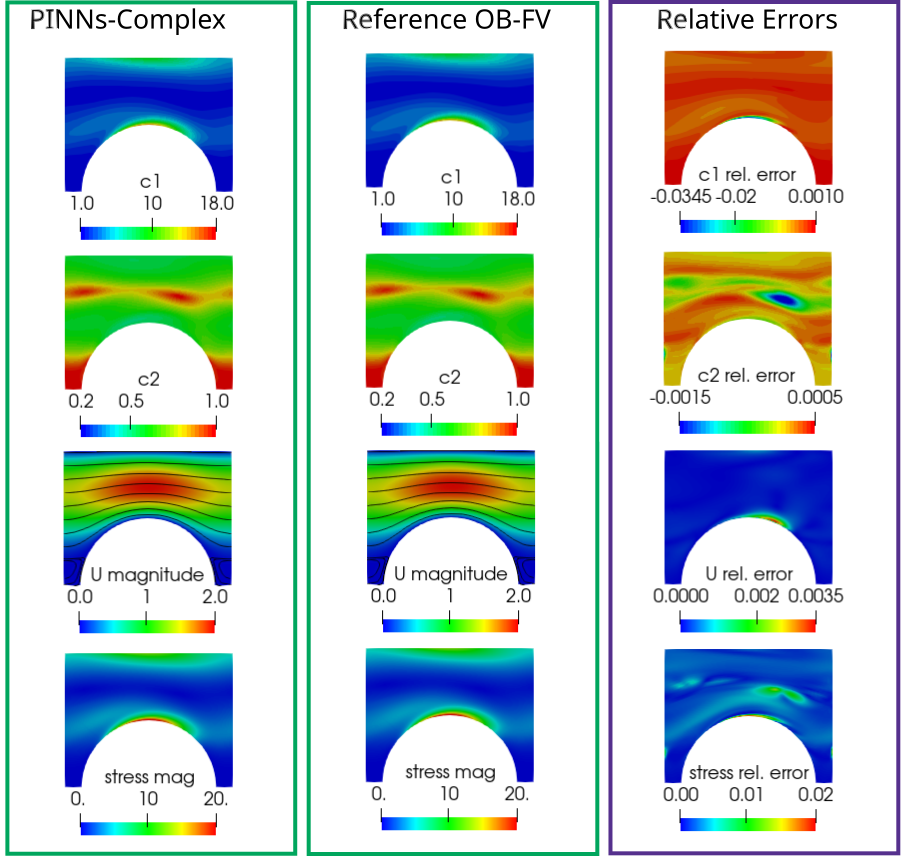}
	\caption{{Comparison of}
 RheoTool simulation with PINN-complex {and} 
 Oldroyd-B models at Wi$=0.4$ in a periodic array of cylinders (PAC) for $L=2.5R$.}
	\label{fig:train_solution_wi0p4_L25}
\end{figure}


The excellent solution of the PINN-complex model verifies the capacity of the model to be geometry agnostic when the explored $c_1$-$c_2$  space  is contained within the training-hull as shown in Fig. \ref{fig:steady_solution_wi0p4_hull_L}}

\clearpage

\section{Conclusions}
We have developed a \textsc{generic} compliant framework that employs PINNs to learn the polymeric contribution to the entropy, which determines the rheological constitutive model. In this approach, the neural network training is guided by a general evolution equation of the conformation tensor.
The model captures the polymeric contribution to the entropy as a function of the eigenvalues of the conformation tensor. This functional approximator can then be easily differentiated using automatic differentiation to compute the stress contributions required in {CFD} simulations.

Our results compare two different methods for training the PINN constitutive models. The first model (PINN-rheometric) is trained with  data generated from the analytical solution of the Oldroyd-B model with the fluid subjected to steady-state rheometric flows. The second model (PINN-complex) is trained with in-silico data generated directly from complex flow  CFD simulations of flow around a cylinder that use Oldroyd-B as a constitutive model. Both PINN models are capable of predicting flow behavior in transient and complex flow conditions that explore regions of the conformational space that are not too far from the domain covered by the training data. However, the PINN-complex model outperforms the PINN-rheometric model in complex flow simulations. {Our results highlights the importance of the regions used to train both models: the wider region used to train the PINN-complex model results in better predictions than the narrow region (a line) used to train the PINN-rheometric model. Furthermore, we apply the PINN-complex model to demonstrate the key advantage of our method being geometry agnostic by performing simulations of flow around an array of cylinders in the conditions in which the flow topology is affected by the hydrodynamic interactions in between cylinders. We are able to capture this phenomenon as well as the underlying Oldroyd-B model does. }
{The PINNs powered modeling framework presented in this manuscript has the potential to change the way constitutive models are created by leveraging data, thermodynamics and physical knowledge and could be in  principle applied to other complex fluids, such as suspensions, provided that a \textsc{generic} framework is available to provide thermodynamically-consistent constraints. }

\section{Acknowledgments}
  This research is supported by the Basque Government through the BERC
  2022-2025 program, the ELKARTEK 2022 and 2024 programs (KAIROS
project: grant KK-2022/00052 and ELASTBAT: KK-2024/00091). The research is also partially funded by the Spanish State Research Agency through BCAM
  Severo  Ochoa excellence  accreditation  CEX2021-0011 42-S/MICIN/AEI/10.13039/501100011033, and  through
  projects   PID2020-117080RB-C55    (``Microscopic   foundations   of
  soft-matter  experiments:   computational  nano-hydrodynamics''  and
  acronym      "Compu-Nano-Hydro"),      and      PID2020-117080RB-C54
  (``Coarse-Graining theory and experimental techniques for multiscale
  biological systems.'') funded by AEI - MICIN.

\bibliographystyle{ieeetr}
\bibliography{aipsamp}

\begin{appendices}
In these appendices, we present several results concerning the OB and FENE models, and analytical results of the OB model for viscometric flows.

\section{The OB and FENE models}

For a system with chains modeled as Hookean dumbbells, the entropy per
unit  volume as  a function  of the  conformation tensor  is given  by
Oldroyd-B model
\begin{align}
  s_{\rm p}(\bm c)
  = \frac{k_{\rm B}}{2}\left({\rm tr}(\bm I - \bm c) + {\rm ln}({\rm det}(\bm c))\right)
  \label{eq:s_OB}
\end{align}
The entropy can be expressed in terms of the eigenvalues of the conformation
tensor $c_1$ and $c_2$,
\begin{equation}
s_{\rm p}(c_1,c_2) = \frac{k_{\rm B}}{2}\left(2 - c_1 - c_2 + \ln{c_1} + \ln{c_2}\right) \label{eq:SOB2}
\end{equation}
If instead the chains are  modeled with finite extensible springs of the FENE type the entropy is
\cite{Ottinger_book}
\begin{align}
s_{\rm p}(\bm c) & = \frac{k_{\rm B}}{2}\left( b \ln \phi(\bm{c}) + \ln \det \bm c\right) \label{eq:s_FENE}\\
    \phi(\bm c) & = \frac{b + D}{b} - \frac{1}{b}{\rm tr}(\bm c) = \frac{b + 2}{b} - \frac{1}{b}(c_1+c_2) \label{eq:phi}
\end{align}
where  dimensionality is given by $D=2$. 
Finally,  $b$  is the  finite
extensibility parameter related to the  springs constant. Note that as \( b \to \infty \), the FENE model converges to the Oldroyd-B model.

The  thermodynamic  force ${\bm  \sigma}$ 
defined in (\ref{eq:sigma_def}) 
for the  Oldroyd-B
model it is given by
\begin{align}
  \label{sigmaOB}
{\bm \sigma} = k_{\rm B}T\left({\bm c}^{-1}-{\bf 1}\right)
\end{align}
while for FENE
\begin{align}
    \bm \sigma = k_{\rm B}T\left(\bm c^{-1} - \frac{\bm I}{\phi(\bm c)}\right)
\end{align}
At equilibrium the entropy is maximal, implying $\bm \sigma=0$. This occurs
in both models for   ${\bf  c}_{\rm  eq}={\bf  1}$. The corresponding entropy
value at equilibrium is zero.

\newpage




\subsection{Steady-State Extensional Flow}\label{sec:ss_ext}

 For steady-sate extensional flow the velocity gradients takes the form in 2D

\begin{align}
\bm \kappa =   \begin{bmatrix} \dot \varepsilon &  0 \\ 0& -\dot \varepsilon \end{bmatrix}
\end{align}
and the conformation tensor is indipendent of space and time and takes the form
 \begin{align}
\bm c   =  \begin{bmatrix} \frac{1}{1-2Wi}&0\\0 &   \frac{1}{1+2Wi} \end{bmatrix}
\end{align}

 In order to construct the residual (\ref{eq:ec_ss}), we need to express the velocity
  gradient in the eigenbasis $\bm u_i = \bm v_i/\| \bm v_i\|$ of the conformation tensor. The components in this basis are

\begin{eqnarray}
\kappa_{\alpha\beta} &\equiv&  {\bm  u}_\alpha^T \!\cdot\!\bm {\kappa} \!\cdot\!{\bm u}_\beta \label{e4}
\end{eqnarray}
%
Note that you can multiply $e_c$ by $\lambda_{\rm p}$ so every thing is given by Wi.  As a result, the $\kappa_{\alpha\beta}$ are given in terms of the Wi number.

\begin{align}\label{eq:kappaext}
\lambda_{\rm p}\kappa_{11}& = Wi \\
\lambda_{\rm p}\kappa_{22}& = -Wi \\
\lambda_{\rm p} \kappa_{12}& = \lambda_{\rm p} \kappa_{21} = 0
\end{align}
and the residual equations become:
\begin{align}\label{eq:res_explicit2}
\tilde{e}^1_c = - 2 c_1 \lambda_{\rm p}\kappa_{11} - c_1 \tilde \sigma_1\\
\tilde{e}^2_c = - 2 c_2 \lambda_{\rm p}\kappa_{22} - c_2 \tilde \sigma_2\\
\end{align}

\subsection{Steady-State Shear Flow}\label{sec:ss_shear}
In this flow the velocity gradient $\bm \kappa$, the conformation tensor $\bm{c}$ and the thermodynamic forces $\bm{\sigma}$ are
\begin{align}
\bm \kappa =   \begin{bmatrix} 0 &  \dot \gamma \\ 0& 0 \end{bmatrix}
\end{align}

	\begin{align*}
     		\bm{c} =   \begin{bmatrix} 1+2\rm{Wi}^2  &  \rm{Wi} \\ \rm Wi & 1 \end{bmatrix}
	\end{align*}
	and
	\begin{align*}
   		 \frac{ \bm \sigma}{k_{\rm B}T} =  \bm c^{-1} - \bm I =   \frac{1}{1+\rm{Wi}^2}\begin{bmatrix} -\rm{Wi}^2  &  -\rm{Wi} \\ -\rm Wi & \rm{Wi}^2 \end{bmatrix}
	\end{align*}
 The eigenvalues and {unnormalized} eigenvectors of the conformation tensor are
	\begin{align}
		c_1 = 1 + \rm{Wi}^2 + \rm{Wi}\sqrt{1 + \rm{Wi}^2} \label{eq:lc1Wi} \\
	    	c_2 = 1 + \rm{Wi}^2 - \rm{Wi}\sqrt{1 + \rm{Wi}^2} \label{eq:lc2Wi}
	\end{align}
%
	\begin{align}\label{eq:shear_vectors}
		     \bm{v}_1 =   \begin{bmatrix} \rm{Wi}+\sqrt{\rm{Wi}^2+1} \\ 1 \end{bmatrix} \\
		     \bm{v}_2 =   \begin{bmatrix} \rm{Wi}-\sqrt{\rm{Wi}^2+1} \\ 1 \end{bmatrix}
		\end{align}

while the eigenvalues of $\bm{\sigma}$ are
\begin{align}
	    	 \frac{\sigma_1}{k_{\rm B}T} = +\frac{\rm{Wi}}{(1+\rm{Wi}^2)^{1/2}} \label{eq:ls1Wi}\\
	       \frac{\sigma_2}{k_{\rm B}T}  = -\frac{\rm{Wi}}{(1+\rm{Wi}^2)^{1/2}} \label{eq:ls2Wi}
	\end{align}

\subsection{Transient solutions to start-up extensional and shear flows}\label{sec:transient}

{\bf Start-up Uniaxial Extension}

The velocity/deformation vector is in general given by $v= (\dot \varepsilon x, -1/2\dot \varepsilon y, -1/2 \dot \varepsilon z)$ for $t>0$. However for our simplified 2D model, we are looking at planar extension. In that case, $v= (\dot \varepsilon x, -\dot \varepsilon y)$  and the velocity gradient is given by:

\begin{align}
    \bm \nabla v = \begin{bmatrix} \dot \varepsilon & 0 \\
    0 & -\dot \varepsilon
    \end{bmatrix}
\end{align}

The evolution of the conformation tensor using the Oldroyd B model for start-up extensional (SUE) flow (i.e., at time $t=0$ stretching begins at strain rate $\dot \varepsilon$):
\begin{align}
    c_{xx} = &\frac{1}{2{\rm Wi}-1} \{2{\rm Wi}\exp{\left[(2{\rm Wi}-1)t/\lambda_{\rm p}\right]} -1\} \label{eq:cxx_ext_tr} \\
    c_{yy} = &\frac{1}{2{\rm Wi}+1} \{2{\rm Wi}\exp{\left[-(2{\rm Wi}+1)t/\lambda_{\rm p}\right]} + 1\} \label{eq:cyy_ext_tr}\\
    c_{xy} = & 0 \label{eq:cxy_ext_tr}
\end{align}
Note that this expressions are limited to Wi$<0.5$. Furthermore, these expressions have the right limits: for $t=0$ we have $c_{xx}= c_{yy} = 1$ and for $t\rightarrow\infty$ we arrive to the steady state extensional flow solution:
\begin{align}
    c_{xx} = & \frac{1}{1-2{\rm Wi}} \\
    c_{yy} = & \frac{1}{1+2{\rm Wi}}
\end{align}

{\bf Start-up Shear Flow}
Oldroyd B start-up shear (SUS) flow (i.e., walls start to move at $t=0$ with shear rate $\dot \gamma = V_{wall}/H$) for $t>0$. The velocity gradient is given by:

\begin{align}\label{eq:theogradv}
            \bm{\nabla v} =   \begin{bmatrix} 0 &  \dot \gamma \\ 0& 0 \end{bmatrix}
\end{align}

\begin{figure}
     \centering
     \begin{subfigure}[b]{0.45\textwidth}
         \centering
         \includegraphics[width=\textwidth]{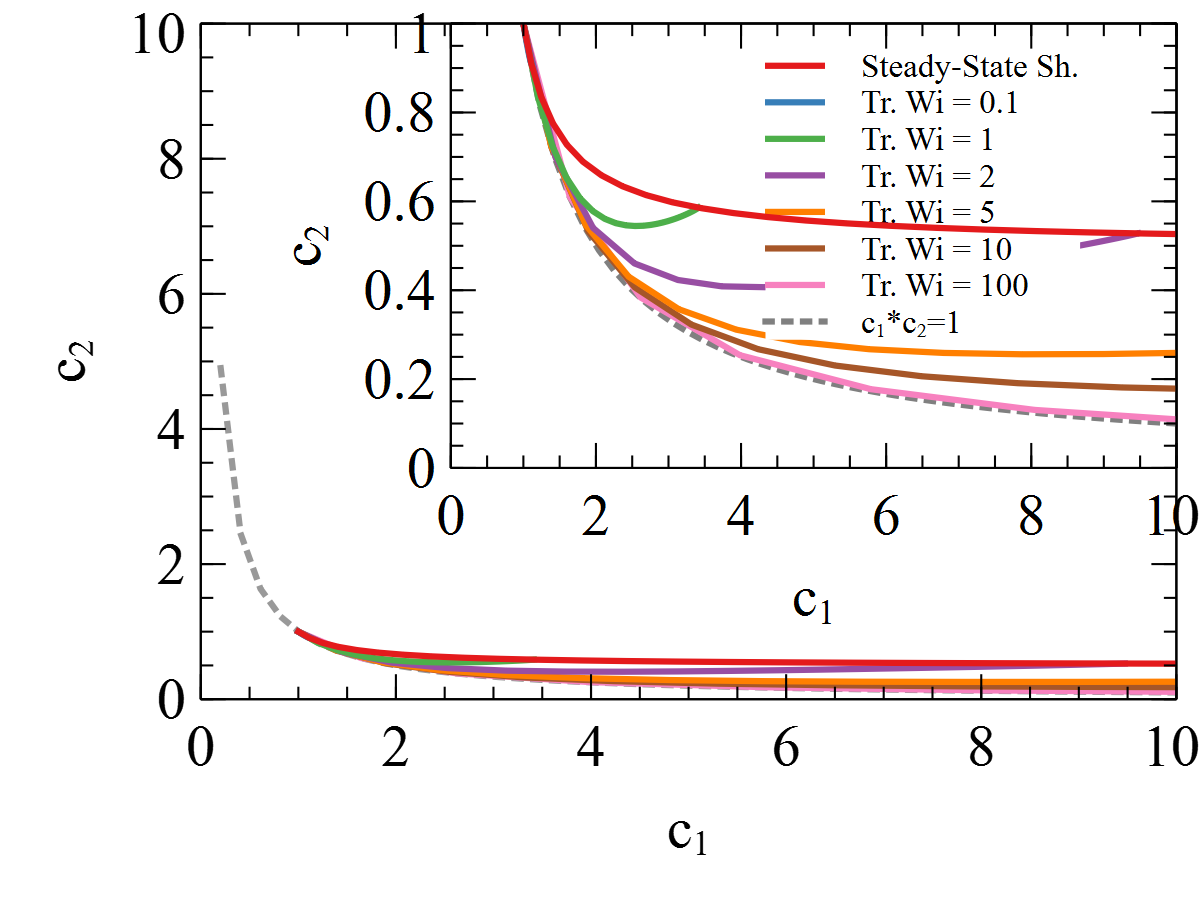}
         \caption{Start-up shear}
         \label{fig:sushear}
     \end{subfigure}
     \hfill
     \begin{subfigure}[b]{0.45\textwidth}
         \centering
         \includegraphics[width=\textwidth]{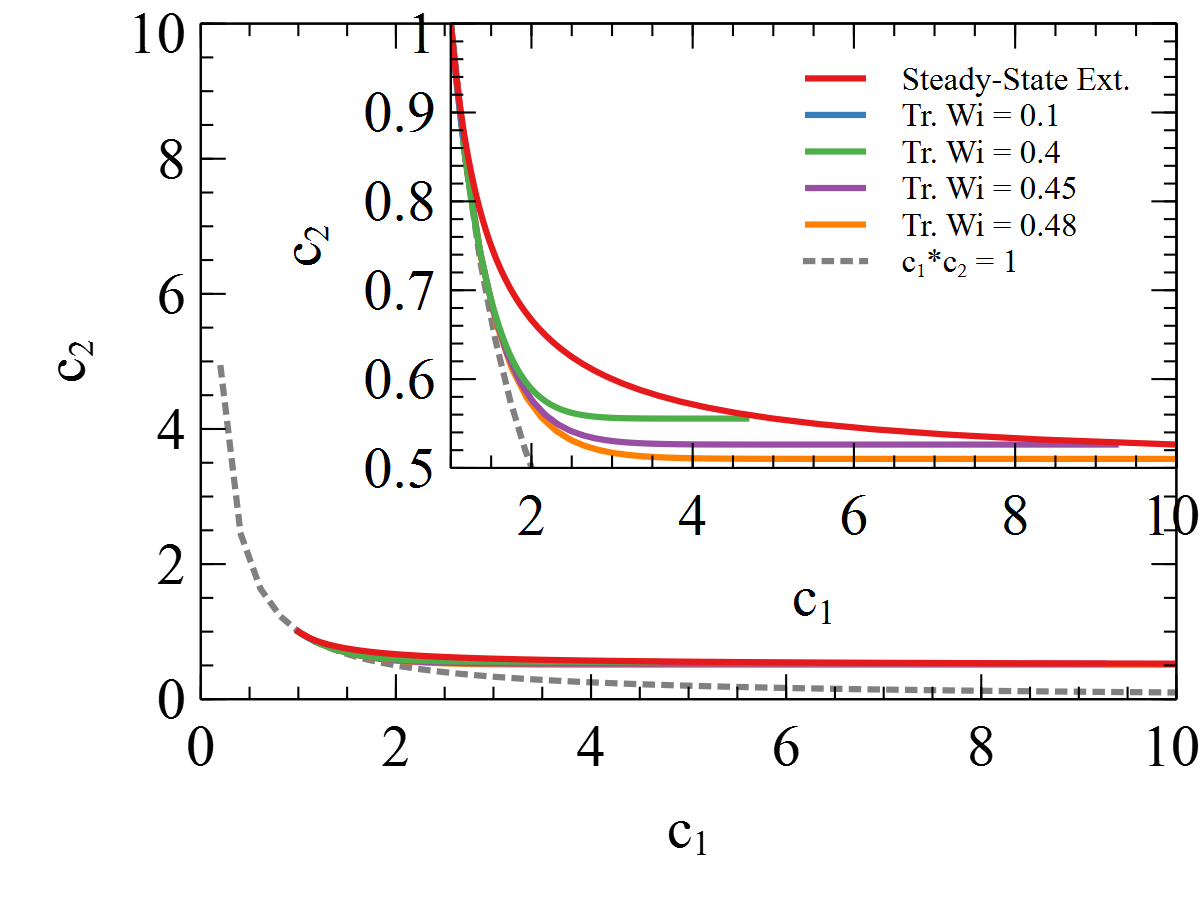}
         \caption{Start-up extensional}
         \label{fig:suext}
     \end{subfigure}
\caption{Startup and steady-state shear and extensional flow analytic solutions for Oldroyd-B model.}
\label{fig:analytic_solutions}
\end{figure}

The analytical solution for the evolution of the conformation tensor is given by
\begin{align}
    c_{xx} = &1 + 2{\rm Wi}^2(1-\frac{t}{\lambda_{\rm p}}\exp\left[-t/\tau_{\rm p}\right] - \exp\left[-t/\lambda_{\rm p}\right]) \label{eq:cxx_she_tr} \\
    c_{yy} = & 1 \label{eq:cyy_she_tr}\\
    c_{xy} = & {\rm Wi}(1- \exp\left[-t/\lambda_{\rm p}\right]) \label{eq:cxy_she_tr}
\end{align}
Here again, the expressions have the right limits for $t=0$ and for $t\rightarrow\infty$ with the steady-state solution:
\begin{align}
    c_{xx} = & 1+2{\rm Wi}^2 \\
    c_{yy} = & 1 \\
    c_{xy} = & Wi
\end{align}

For shear flow:
\begin{align}\label{eq:kappashear}
\lambda_{\rm p}\kappa_{11}= \frac{{\rm Wi}^2 + {\rm Wi}\sqrt{{\rm Wi}^2 +1}}{\| v_1\|^2} \\
\lambda_{\rm p}\kappa_{22}= \frac{{\rm Wi}^2 - {\rm Wi}\sqrt{{\rm Wi}^2 +1}}{\| v_2\|^2} \\
\lambda_{\rm p}\kappa_{12} =  \frac{{\rm Wi}^2 - {\rm Wi}\sqrt{{\rm Wi}^2 +1}}{\| v_1\| \| v_2\|} \\
\lambda_{\rm p}\kappa_{21} =  \frac{{\rm Wi}^2 + {\rm Wi}\sqrt{{\rm Wi}^2 +1}}{\| v_1\| \| v_2\|}
\end{align}

where we have used $\bm u_i = \bm v_i/\| v_i\|$ , where $\bm v_i$ are the eigenvectors in  (\ref{eq:shear_vectors})  and:
\begin{align}\label{eq:umod}
\| v_1\|= \sqrt{2[1+{\rm Wi}^2 + {\rm Wi}\sqrt{{\rm Wi}^2 +1}]} \\
\| v_2\|= \sqrt{2[1+{\rm Wi}^2 - {\rm Wi}\sqrt{{\rm Wi}^2 +1}]}
\end{align}

The analytic solutions for both shear and extensional steady-state and start-up flows are shown in Fig. \ref{fig:analytic_solutions}

\end{appendices}

\end{document}